%% file: main.tex
\documentclass[conference,10pt,letterpaper,nofonttune]{IEEEtran}
\pdfoutput=1
\IEEEoverridecommandlockouts
\usepackage{cite}
\usepackage{csquotes}
\usepackage{amsmath,amssymb,amsfonts}
\usepackage{algorithmic,algorithm}
\usepackage{xcolor}
\usepackage{subfigure}
\usepackage{comment}
\usepackage{xspace}
\usepackage{balance}
\usepackage{fancyhdr}

\newcommand{\mycomment}[1]{}
\newcommand{\oran}{O-RAN\xspace}

\usepackage{soul}
\usepackage{multirow}
\usepackage{booktabs}
\usepackage{tabularx}
\usepackage{tikz}
\usepackage{pgfplots}
\usepackage{adjustbox}
\pgfplotsset{compat=newest}
\pgfplotsset{plot coordinates/math parser=false}
\newlength\fheight
\newlength\fwidth
\usetikzlibrary{plotmarks,patterns,decorations.pathreplacing,backgrounds,calc,arrows,arrows.meta,spy,matrix,scopes}
\usepgfplotslibrary{patchplots,groupplots}
\usepackage{tikzscale}
\usepackage{hyperref}

\newif\ifexttikz
\exttikzfalse

\ifexttikz
	\usetikzlibrary{external}
	\usepackage{fontspec}
\fi

\usepackage[acronyms,nonumberlist,nopostdot,nomain,nogroupskip,acronymlists={hidden}]{glossaries}
\newglossary[algh]{hidden}{acrh}{acnh}{Hidden Acronyms}
\input{acronyms.tex}

\glsdisablehyper

\fancypagestyle{FirstPage}{
    \lhead{\small{This work has been accepted for publication in IEEE GLOBECOM 2023.} \\
    \scriptsize{\copyright{} 2023 IEEE. Personal use of this material is permitted. Permission from IEEE must be obtained for all other uses, in any current or future media,
    including reprinting/republishing this material for advertising or promotional purposes, creating new collective works, for resale or 
    redistribution to servers or lists, or reuse of any copyrighted component of this work in other works.}}
}

\begin{document}

\title{A Comparative Analysis of Deep Reinforcement Learning-based xApps in O-RAN\vspace{-.15cm}}

\author{\IEEEauthorblockN{Maria Tsampazi\IEEEauthorrefmark{1}, Salvatore D'Oro\IEEEauthorrefmark{1}, Michele Polese\IEEEauthorrefmark{1}, Leonardo Bonati\IEEEauthorrefmark{1},\\Gwenael Poitau\IEEEauthorrefmark{4}, Michael Healy\IEEEauthorrefmark{4}, Tommaso Melodia\IEEEauthorrefmark{1}}
\IEEEauthorblockA{\IEEEauthorrefmark{1}Institute for the Wireless Internet of Things, Northeastern University, Boston, MA, U.S.A.\\E-mail: \{tsampazi.m, s.doro, m.polese, l.bonati, melodia\}@northeastern.edu\\\IEEEauthorrefmark{4}Dell Technologies, P\&O OCTO – Advanced Wireless Technology\\E-mail: \{gwenael.poitau, mike.healy\}@dell.com}

\thanks{This article is based upon work partially supported by Dell Technologies and by the U.S.\ National Science Foundation under grants CNS-1925601, CNS-2112471, CNS-1923789 and CNS-2120447.}
}

\makeatletter
\patchcmd{\@maketitle}
  {\addvspace{0.5\baselineskip}\egroup} 
  {\addvspace{-1.5\baselineskip}\egroup} 
  {}
  {}
\makeatother

\maketitle

\glsunset{usrp}

\begin{abstract}
The highly heterogeneous ecosystem of \gls{nextg} wireless communication systems calls for novel networking paradigms where functionalities and operations can be dynamically and optimally reconfigured in real time to adapt to changing traffic conditions and satisfy stringent and diverse \gls{qos} demands.
Open \gls{ran} technologies, and specifically those being standardized by the \oran Alliance, make it possible to integrate network intelligence into the once monolithic \gls{ran} via intelligent applications, namely, xApps and rApps.
These applications enable flexible control of the network resources and functionalities, network management, and orchestration
through data-driven control loops. Despite recent work demonstrating the effectiveness of \gls{drl} in controlling \oran systems, how to
design these solutions in a way that does not create conflicts and unfair resource allocation policies is still an open challenge.
%
In this paper, we perform a comparative analysis where we dissect the impact of different \gls{drl}-based xApp designs on network performance. Specifically, we benchmark 12 different xApps that embed \gls{drl} agents trained using different reward functions, with different action spaces and with the ability to hierarchically control different network parameters. 
%
We prototype and evaluate these xApps on Colosseum, the world's largest \oran-compliant wireless network emulator with hardware-in-the-loop. 
We share the lessons learned and discuss our experimental results, which demonstrate how certain design choices deliver the highest performance while others might result in a competitive behavior between different classes of traffic with similar objectives.

\end{abstract}

\glsresetall
\glsunset{usrp}


\section{Introduction} \label{Section I}
\thispagestyle{FirstPage}
Programmable, 
virtualized, and disaggregated architectures are seen as key enablers of \gls{nextg} cellular networks. Indeed, the flexibility offered through softwarization, virtualization, and open standardized interfaces provides new self-optimization capabilities. These concepts are at the foundation of the Open \gls{ran} paradigm, which is being specified by the \oran Alliance. Thanks to the \glspl{ric} proposed by \oran (i.e., the near- and non-real-time \glspl{ric}), intelligence can be embedded into the network and leveraged for on-demand closed-loop control of its resources and functionalities~\cite{polese2023understanding}. In \oran, this is achieved via intelligent applications, called xApps and rApps, which execute on the near- or non-real-time \glspl{ric} respectively.
Through the \glspl{ric}, these applications interface with the network nodes and implement data-driven closed-loop control based on real-time statistics received from the \gls{ran}, thus realizing the vision of resilient, reconfigurable and autonomous networks~\cite{polese2023understanding}.
%
Since they do not require prior knowledge of the underlying network dynamics~\cite{sutton2018reinforcement}, \gls{drl} techniques are usually preferred in the design of such control solutions for the Open \gls{ran}~\cite{wang2022self,polese2022colo,d2022orchestran}.


Intelligent control in \oran through xApps has widely attracted the interest of the research community. For example, \cite{johnson2022nexran} proposes the NexRAN xApp to control
and balance the throughput of different \gls{ran} slices.
The authors of~\cite{kouchaki2022actor} develop a \gls{rl} xApp to assign resource blocks to certain users according to their \gls{csi} and with the goal of maximizing the aggregated data rate of the network.
A deep Q-learning-based xApp for controlling slicing policies to minimize latency for \gls{urllc} slices is presented in \cite{filali2023communication}.
The authors of~\cite{polese2022colo} experimentally evaluate and demonstrate several \gls{drl}-based xApps under a variety of traffic and channel conditions, and investigate how different action space configurations impact the network performance. 
Finally, other research efforts focus on coordinating multiple xApps to control different parameters via a combination of federated learning and team learning~\cite{iturria2022multi,zhang2022federated}.


\subsection{Contributions and Outline} \label{Section IB}

The above works clearly show that \gls{drl} and \gls{ai},
are catalysts in the design and development of intelligent control solutions for the Open \gls{ran}. However, despite early results showing their success and effectiveness, designing \gls{drl} agents for complex Open \gls{ran} scenarios---characterized by the coexistence of diverse traffic profiles and potentially conflicting \gls{qos} demands---is still an open challenge that, as we describe below, we aim at addressing in this paper. Specifically, our goal is to go beyond merely using \gls{ai}, and specifically \gls{drl}, in a black-box manner. Instead, we try to address some fundamental questions that are key for the success of intelligence in Open \gls{ran} systems.

We consider an Open \gls{ran} delivering services to \gls{urllc}, \gls{mmtc} and \gls{embb} network slices.
Specifically, we use OpenRAN Gym~\cite{bonati2023openran}---an open-source framework for \gls{ml} experimentation in \oran---to deploy such Open \gls{ran} on the Colosseum wireless network emulator~\cite{bonati2021colosseum}, and control it through 12~\gls{drl} xApps. 
%
%
These xApps have been trained to perform slice-based resource allocation (i.e., scheduling profile selection and \gls{ran} slicing control) and to meet the diverse requirements of each slice.
We investigate the trade-off between long-term and short-term rewards, we discuss and compare different design choices of action set space, hierarchical decision-making policies and action-taking timescales.
Finally, we show how these choices greatly impact network performance and affect each slice differently.  
 
To the best of our knowledge, this is the first experimental study that provides a comprehensive evaluation of the design choices for \gls{drl}-based xApps. We hope that our findings and insights might help in designing xApps for \gls{nextg} Open \glspl{ran}.


The remainder of this paper is organized as follows.
Section~\ref{Section II} describes our system model and data-driven optimization framework.
Section~\ref{Section III} presents the different \gls{drl} optimization strategies considered in this work, while Section~\ref{Section IV} details our experimental setup and training methodology.
Experimental results are discussed in Section~\ref{sec:experimental-evaluation}.
Finally, Section~\ref{Section V} draws our conclusions and presents some future work directions.

\section{System Model and Data-Driven Optimization Framework} \label{Section II}

In this work, we consider an Open \gls{ran} multi-slice scenario where \glspl{ue} generate traffic with diverse profiles and \gls{qos} demands. Without loss of generality, we assume that traffic generated by \glspl{ue} can be classified into \gls{embb}, \gls{urllc}, or \gls{mmtc} slices.  

\begin{figure}[t!]
  \centering
  \includegraphics[width=3.5in]{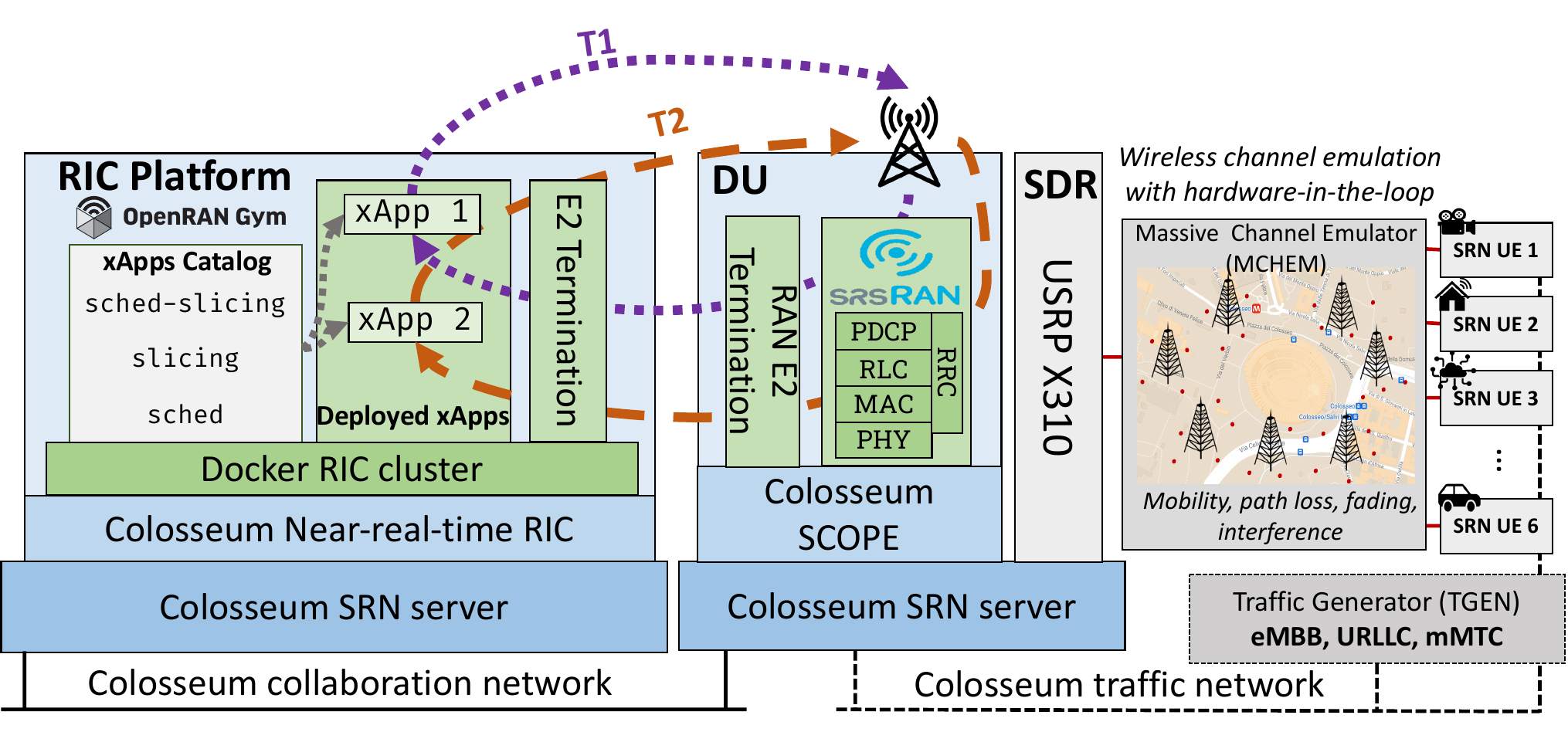} 
  \setlength\abovecaptionskip{-.1cm}
  \caption{Reference \oran testing architecture with focus on the case of two xApps operating at different time scales, $T_{i}$, as described in Section \ref{Section IV-B}.}
  \label{fig:ext-arch}
  \vspace{-0.45cm}
\end{figure}

To satisfy the diverse \gls{qos} demands required by each slice and intelligently control the resource allocation process, we leverage \gls{rl}. Specifically, as shown in Fig.~\ref{fig:ext-arch}, we leverage xApps embedding \gls{drl} agents that are tasked with reconfiguring control parameters of the \gls{bs} such as \gls{ran} slicing (i.e., the portion of available \glspl{prb} that are allocated to each slice at any given time) and \gls{mac} layer scheduling policies (in our case, by selecting a dedicated scheduler profile to each slice among \gls{rr}, \gls{wf} and \gls{pf}). xApps make decisions based on the \glspl{ue} traffic demand, load, performance and network conditions that are given by \glspl{kpm} periodically reported by \gls{ran}. 


\subsection{DRL Agent Architecture}

We focus on \gls{drl} agents that implement the \gls{ppo} algorithm, a state-of-the-art on-policy \gls{drl} algorithm based on an actor-critic network architectural approach. Specifically, the actor and critic network \blockquote{work} cooperatively to learn a policy that selects actions that deliver the highest reward possible for each state. While the actor's task is to take actions based on current network states, the critic's target is to evaluate actions taken by the actor network and provide feedback that reflects how effective the action taken by the actor is. In this way, the critic helps the actor in taking actions that lead to the highest rewards for each given state.

The reason we focus on \gls{ppo} is that it has been demonstrated several times to outperform other architectures~\cite{polese2022colo,kouchaki2022actor}. 
Actor and critic networks are fully-connected neural networks with 3 layers of 30 neurons each. The hyperbolic tangent serves as the activation function while the learning rate is set to $10^{-3}$. We follow the same approach as in \cite{polese2022colo}, where the input (e.g., \glspl{kpm}) of the \gls{drl} agent is first processed by the encoding part of an autoencoder for dimensionality reduction. This also synthetically reduces the state space and makes training more efficient in terms of time and generalization. In detail, the autoencoder converts an input matrix of $K=10$ individual measurements of $M=3$ \gls{kpm} metrics (i.e., downlink throughput, buffer occupancy, and number of transmitted packets) into a single $M$-dimensional vector. The \gls{relu} activation function and four fully-connected layers of $256$, $128$, $32$ and $3$ neurons are also used in the encoder.

The cumulative average reward function of the \gls{drl} agent is designed to jointly satisfy the \gls{qos} demand of the three slices with respect to their \gls{kpm} requirements. For instance, \gls{embb} users aim to maximize throughput, while \gls{mmtc} users aim at maximizing the number of transmitted packets. Finally, the goal of \gls{urllc} users is to deliver packets with minimum latency. Since the base station cannot measure end-to-end application-layer latency (which is instead measured at the receiver side), we measure latency in terms of number of bytes in the transmission buffer, the smaller the buffer, the smaller the latency.
The reward is
formulated as the weighted sum in Eq.~\eqref{eq:weighted_reward}

{\small
\begin{equation}\label{eq:weighted_reward}
   R = \sum_{t=0}^{\infty} \gamma^t \left( \sum_{j=1}^{N} w_{j} \cdot r_{j,t} \right), 
\end{equation}%
}%
where $t$ represents the training step, and $N=3$ is the total number of slices, $w_{j}$ represents the weight associated to slice $j$, considered for reward maximization in the three corresponding slices. Finally, $\gamma$ is the discount factor and $r_{j,t}$ describes the slice-specific reward obtained at each training step $t$. In our case, $r_{j,t}$ represents the average value of the \gls{kpm} measured by all users of slice $j$ at time $t$ (e.g., throughput for the \gls{embb} slice). Note that the weight $w_{j}$ for the \gls{urllc} slice is negative to model the minimization of the buffer occupancy. The models that we have designed and trained are deployed as xApps on the near-real-time \gls{ric}, as illustrated in Fig.~\ref{fig:ext-arch}.

\section{DRL Optimization Strategies} \label{Section III}

We investigate how different design choices affect the effectiveness and decision-making
of the \gls{drl}-based xApps.
We consider the following design choices, for a total of 12 xApps.

\textbf{Short-term vs. Long-term Rewards.}
\label{Section IIIA}
We train \gls{drl} agents with different values of the discount factor $\gamma$. The \gls{ppo} discount factor weights instantaneous rewards against long-term rewards. A higher value prioritizes long-term rewards, while a lower $\gamma$ prioritizes short-term rewards. Results of this exploration are provided in Section \ref{Section IV-A}.

\textbf{Hierarchical Decision-Making.} \label{Section IIIB}
We investigate the case of two xApps configuring different parameters in parallel but at different timescales. In this way, we investigate how multiple xApps with different optimization goals and operating timescales impact the overall network performance. The findings of this investigation are provided in Section \ref{Section IV-B}, and
a practical example is illustrated in Fig.~\ref{fig:ext-arch}.

\textbf{Impact of Reward's Weights.} \label{Section IIIC}
Finally, we test different values for the weights $w_i$ of the slices in Eq.~\eqref{eq:weighted_reward}. A different weight configuration affects how \gls{drl} agents prioritize each slice. The results of this analysis are reported in Section \ref{Section IV-C}, where we show how weights significantly impact the overall performance and can result in inefficient control strategies.


\section{Experimental Setup and DRL Training}
\label{Section IV}

\begin{figure*}[t!]
\centering
\subfigure[\gls{embb} Throughput]{\includegraphics[height=2.85cm]{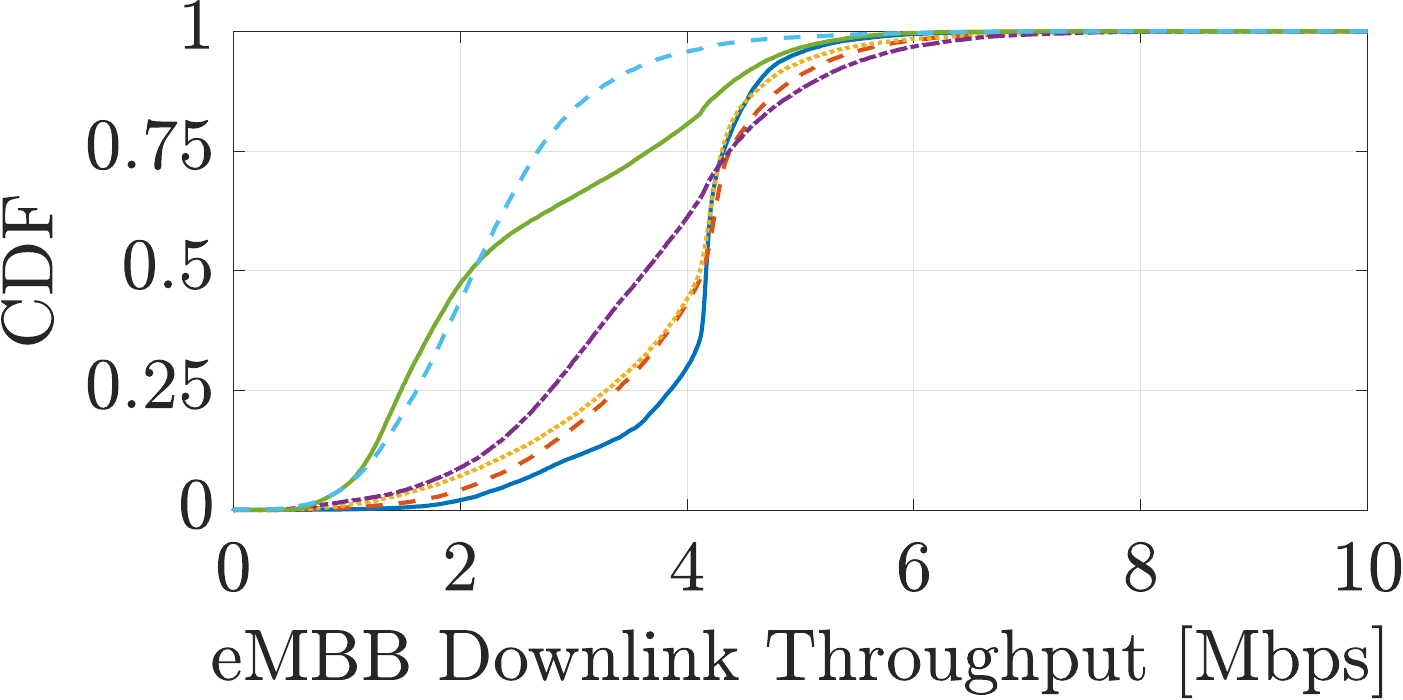}
\label{fig:Figure4a}}
\hfil
\subfigure[\gls{mmtc} Packets]{\includegraphics[height=2.85cm]{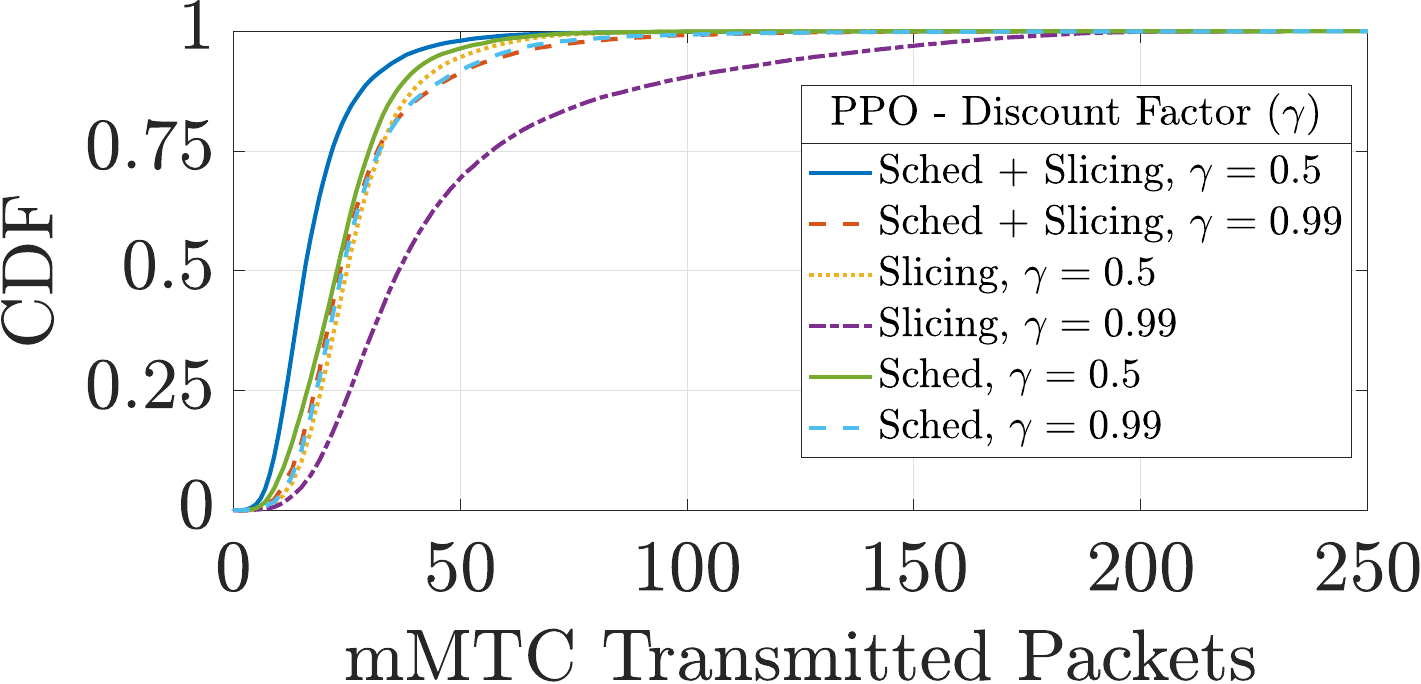}
\label{fig:Figure4b}}
\hfil
\subfigure[\gls{urllc} Buffer Occupancy]{\includegraphics[height=2.85cm]{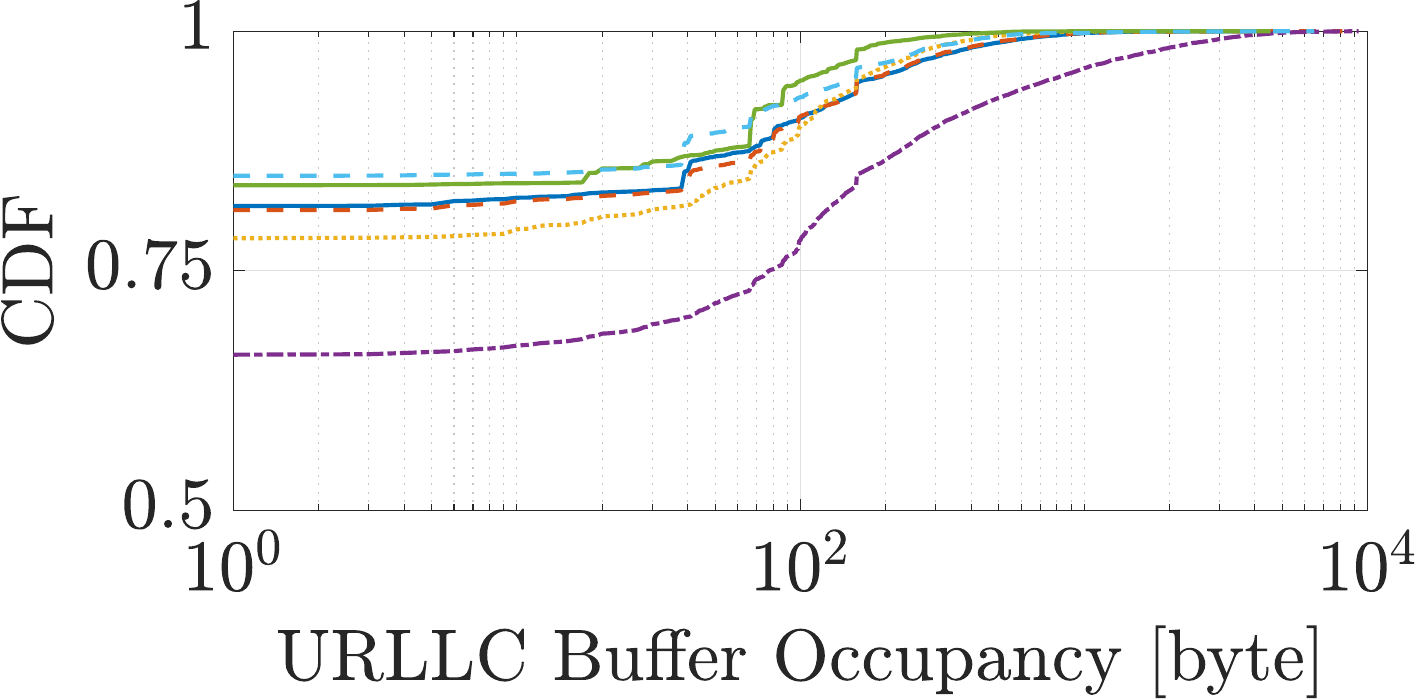}
\label{fig:Figure4c}}
\setlength\abovecaptionskip{-.02cm}
\caption{Performance evaluation under different action spaces and values of the $\gamma$ parameter.}
\label{Figure4-1a}
\vspace{-0.55cm}
\end{figure*}

\begin{figure}[t!]
\centering
\subfigure[\gls{embb} DL Throughput]{\includegraphics[width=1.6in]{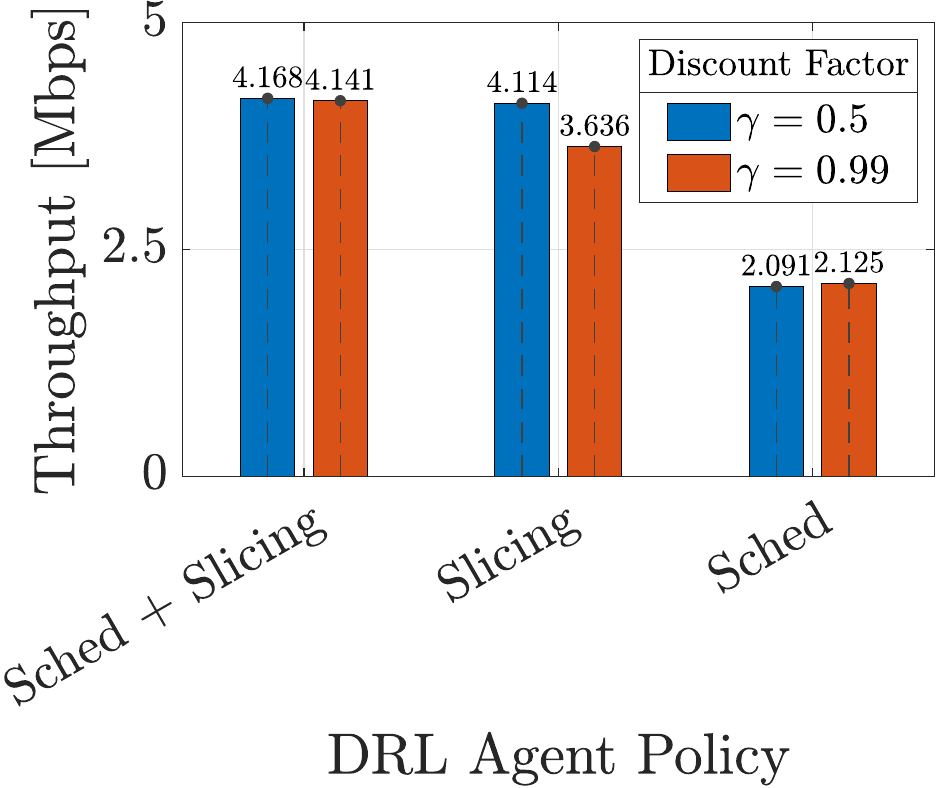}
\label{fig:Figure4d}}
\hfil
\subfigure[\gls{mmtc}  Packets]{\includegraphics[width=1.6in]{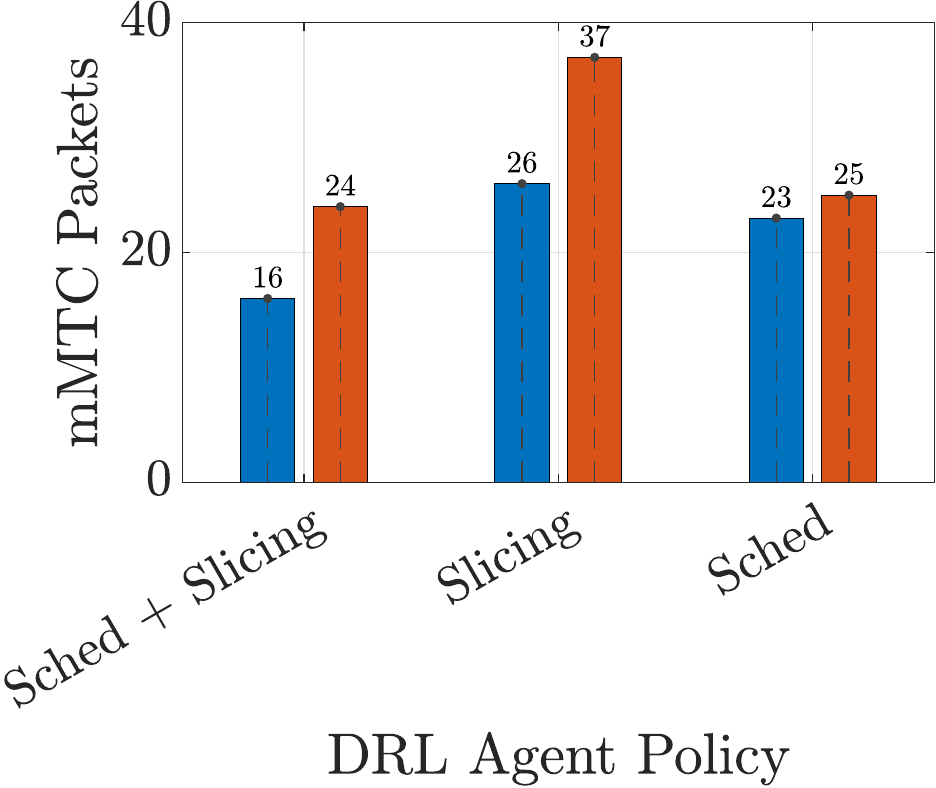}
\label{fig:Figure4e}}
\setlength\abovecaptionskip{-.02cm}
\caption{Median values under different action spaces and values of $\gamma$.}
\label{Figure4-1b}
\end{figure}

To experimentally evaluate the \gls{drl} agents, we leverage the capabilities of OpenRAN Gym \cite{bonati2023openran}, an open-source experimental toolbox for end-to-end design, implementation, and testing of \gls{ai}/\gls{ml} applications in \oran. It features:
\begin{itemize}
   \item End-to-end \gls{ran} and core network deployments though the srsRAN~\cite{gomez2016srslte} softwarized open-source protocol stack; 

    \item Large-scale data collection, testing and fine-tuning of \gls{ran} functionalities through the SCOPE framework~\cite{bonati2021scope}, which adds open \glspl{api} to srsRAN for the control of slicing and scheduling functionalities, as well as for \glspl{kpm} collection;
    
    \item An \oran-compliant control architecture to execute \gls{ai}/\gls{ml}-based xApps via the ColO-RAN near-real-time \gls{ric}~\cite{polese2022colo}. 
    The E2 interface between \gls{ran} and the \gls{ric} and its \glspl{sm}~\cite{polese2023understanding} manage streaming of \glspl{kpm} from the RAN and control actions from the xApps. 
\end{itemize}

We deploy OpenRAN Gym on Colosseum~\cite{bonati2021colosseum}, a publicly available testbed with $128$~\glspl{srn}, i.e., pairs of Dell PowerEdge R730 servers and NI \gls{usrp} X310 \glspl{sdr}.
Colosseum enables large-scale experimentation in diverse \gls{rf} environments and network deployments.
This is done through the \gls{mchem} component, which leverages \gls{fpga}-based \gls{fir} filters to reproduce different conditions of the wireless environment modeled a priori through ray-tracing software, analytical models, or real-world measurements.
The channel conditions that can be emulated in this way include  path loss, fading, attenuation, mobility and interference of signals.
Similarly, the Colosseum \gls{tgen}, built on top of the \gls{mgen}
TCP/UDP traffic generator \cite{mgen}, emulates different traffic profiles (e.g., multimedia content), demand, and distributions (e.g., Poisson, periodic).
%



We deploy a 3GPP-compliant cellular network with one base station and $6$~\glspl{ue} uniformly distributed across $3$~different slices.
These are: (i)~\gls{embb} that concerns high traffic modeling of high-quality multimedia content and streaming applications; (ii)~\gls{urllc} for time-critical applications, such as autonomous driving in \gls{v2x} scenarios; and (iii)~\gls{mmtc} for \gls{iot} devices with low data rate requirements
but with high need for consistent information exchange.
In terms of physical deployment, \glspl{ue} are uniformly distributed within a $20$\:m radius from the \gls{bs}, in the urban environment of Rome, Italy~\cite{bonati2021scope}.

The bandwidth of the \gls{bs} is set to $10$\:MHz (i.e., $50$ \glspl{prb}) and is divided among the $3$ slices, with $2$ users statically assigned to each slice.
Slice-based traffic is created with the following specifications: \gls{embb} users request $4$\:Mbps constant bitrate, while \gls{urllc} and \gls{mmtc} \glspl{ue} generate $89.3$\:kbps and $44.6$\:kbps Poisson traffic, respectively.

To train the \gls{drl} agents, we used the publicly available dataset described in~\cite{polese2022colo}.
This dataset contains about $8$\:GB of \glspl{kpm} collected by using OpenRAN Gym and the Colosseum network emulator over $89$~hours of experiments, and concerns
setups with up to 7~base stations and 42~\glspl{ue} belonging to different \gls{qos} classes, and served with heterogeneous scheduling policies.
%
Each \gls{drl} model evaluated in the following sections takes as input \gls{ran} \glspl{kpm} such as throughput, buffer occupancy, number of \glspl{prb}, and outputs resource allocation policies (e.g., RAN slicing and/or scheduling) for \gls{ran} control. 

Abiding by the \oran specifications, we train our \gls{ml} model offline on Colosseum's GPU-accelerated environment, which includes two NVIDIA DGX A100 servers with $8$~GPUs each. Trained \gls{drl}-agents are onboarded on xApps inside softwarized containers implemented via Docker and deployed on the ColO-RAN near-real-time \gls{ric}. 

\begin{figure*}[t!]
\centering
\subfigure[\gls{embb} Throughput]{\includegraphics[height=2.85cm]{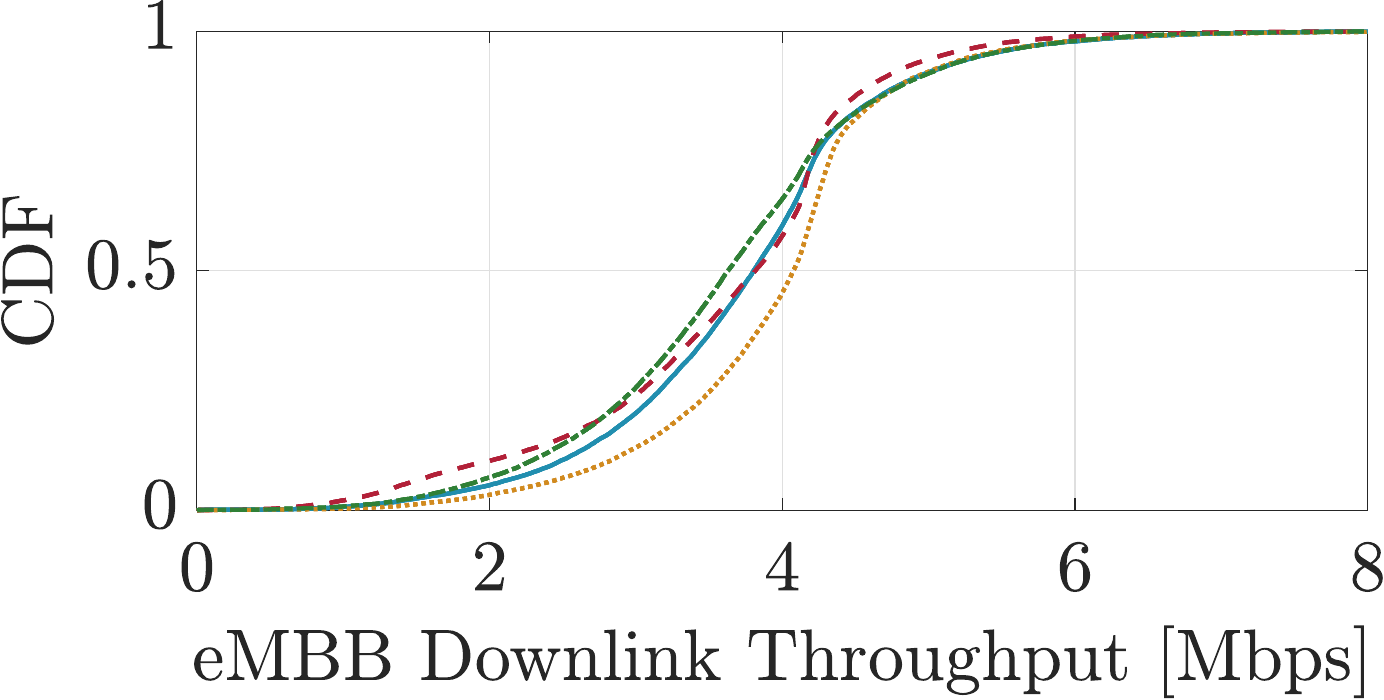}
\label{Figure4a2}}
\hfil
\subfigure[\gls{mmtc} Packets]{\includegraphics[height=2.85cm]{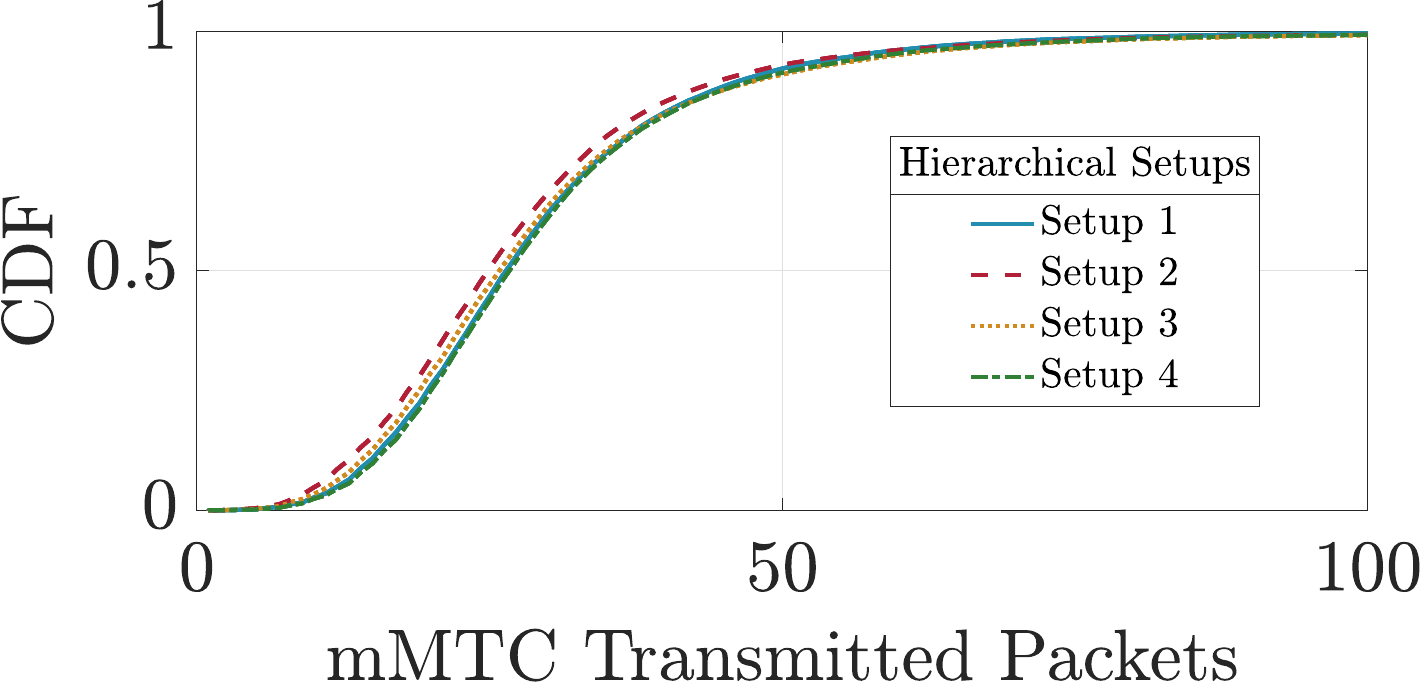}
\label{Figure4b2}}
\hfil
\subfigure[\gls{urllc} Buffer Occupancy]{\includegraphics[height=2.85cm]{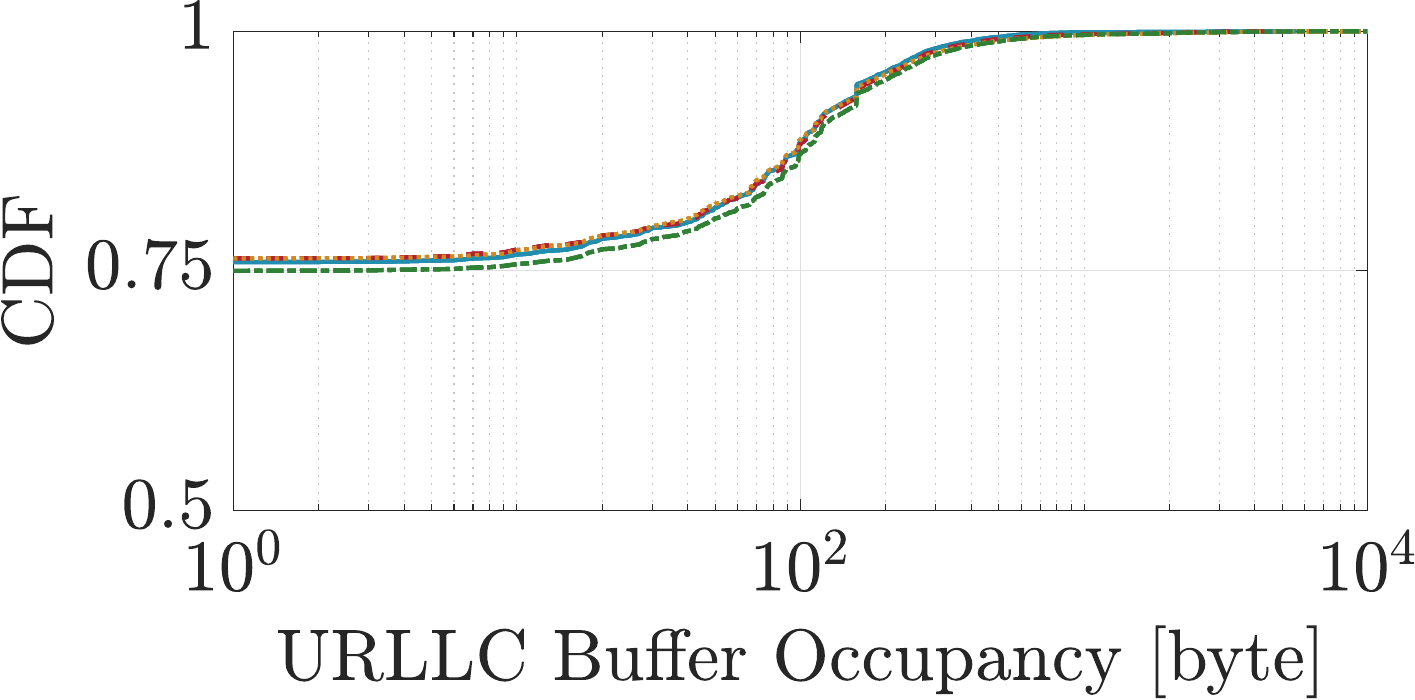}
\label{Figure4c2}}
\setlength\abovecaptionskip{-.02cm}
\caption{Performance evaluation under different hierarchical configurations.}
\label{Figure4-2a}
\vspace{-0.55cm}
\end{figure*}

\begin{figure}[t!]
\centering
\subfigure[\gls{embb} DL Throughput]{\includegraphics[width=1.6in]{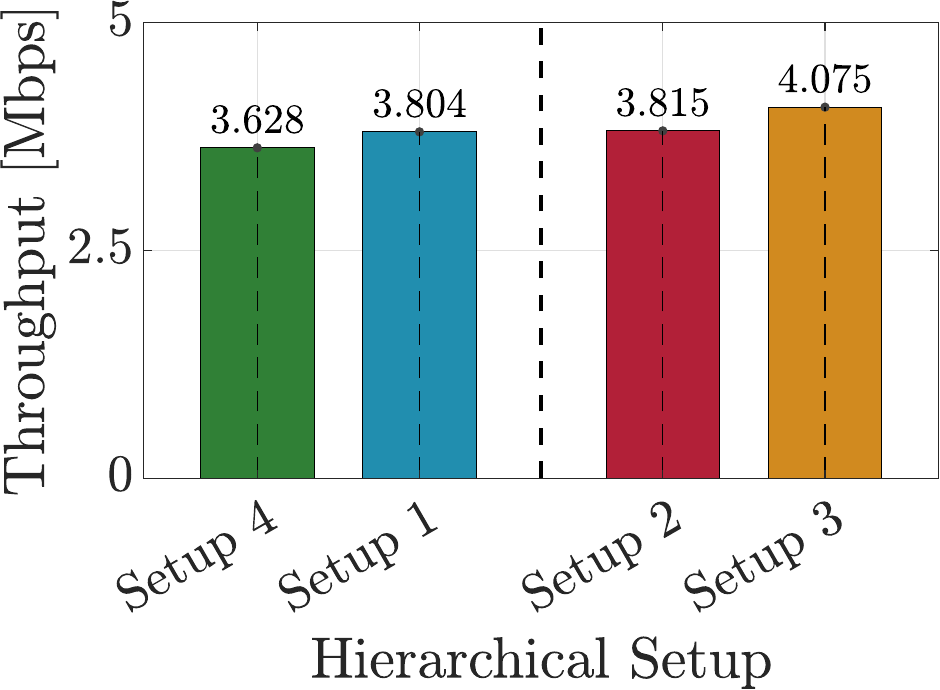}
\label{Figure4d2}}
\hfil
\subfigure[\gls{mmtc}  Packets]{\includegraphics[width=1.6in]{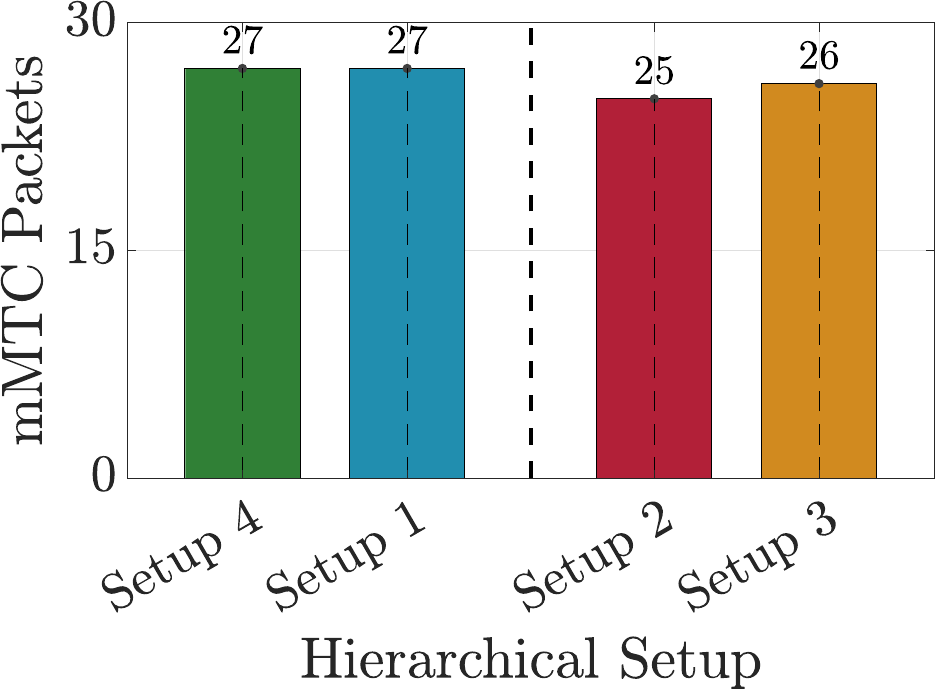}
\label{Figure4e2}}
\setlength\abovecaptionskip{-.02cm}
\caption{Median values under different hierarchical configurations.}
\label{Figure4-2b}
\end{figure}

\section{Experimental Evaluation}
\label{sec:experimental-evaluation}

In this section, we present the results of an extensive performance evaluation campaign, with more than $16$~hours of experiments, to profile the impact of the strategies discussed in Section~\ref{Section III}. These results were produced by taking the median as the most representative statistical value of a dataset, and averaged over multiple repetitions of experiments in the \gls{dl} direction of the communication system.

\subsection{Impact of Discount Factor on the Action Space}\label{Section IV-A}

We explore how \gls{ran} slicing, \gls{mac} scheduling, and joint slicing and scheduling control are affected by training procedures that favor short-term against long-term rewards. Due to space limitations, we only report results obtained with $\gamma\in\{0.5,0.99\}$.
The reward's weight configuration used in this study is shown in Table \ref{table:weight-confs-list} and identified as \texttt{Default}.

\begin{table}[bt]
\centering
\small
\setlength\abovecaptionskip{-.1cm}
\caption{Weight Configurations}
\begin{adjustbox}{width=0.75\linewidth}
\begin{tabular}{@{}l@{\hspace{3mm}}l@{\hspace{3mm}}l@{\hspace{3mm}}l@{}}
\toprule
\multicolumn{1}{c}{\textbf{Weights}} & \multicolumn{1}{c}{\textbf{eMBB}} & \multicolumn{1}{c}{\textbf{mMTC}}  & \multicolumn{1}{c}{\textbf{URLLC}} \\ \midrule 
\texttt{Default} & $72.0440333$ & $0.229357798$ & $0.00005$ \\
\texttt{Alternative} & $72.0440333$ & $1.5$ & $0.00005$ \\
\bottomrule
\end{tabular}
\end{adjustbox}
\label{table:weight-confs-list}
\vspace{-.3cm}
\end{table}

In Fig.~\ref{Figure4-1a}, we report the \gls{cdf} of individual \glspl{kpm} for each slice and for different xApps trained to control different sets of actions and using different values of $\gamma$. The median of such measurements for the \gls{embb} and \gls{mmtc} slices is instead reported in Fig.~\ref{Figure4-1b}.
The median for the \gls{urllc} slice is not reported, as this value is zero in all configurations.
The best performing configurations for the \gls{embb} and \gls{mmtc} slices are instead listed in numerical order in Table~\ref{table:default-comp-analysis} from best to worst performing.

\begin{table}[ht]
\centering
\small
\setlength\abovecaptionskip{-.1cm}
\caption{Per-Slice Top performing xApps under the Default Weight Configuration}
\begin{adjustbox}{width=0.85\linewidth}
\begin{tabular}{@{}l@{\hspace{0.25mm}}ll@{\hspace{0.25mm}}l@{}}
\toprule
\multicolumn{1}{c}{\textbf{}} & \multicolumn{1}{c}{\textbf{eMBB}} & \multicolumn{1}{c}{\textbf{mMTC}} \\ \midrule
\texttt{1)} & \texttt{Sched \& Slicing 0.5} & \texttt{Slicing 0.99}  \\
\texttt{2)} & \texttt{Sched \& Slicing 0.99} & \texttt{Slicing 0.5}  \\
\texttt{3)} & \texttt{Slicing 0.5} & \texttt{Sched 0.99} \\
\texttt{4)} & \texttt{Slicing 0.99} & \texttt{Sched \& Slicing 0.99}  \\
\texttt{5)} & \texttt{Sched 0.99} & \texttt{Sched 0.5}  \\
\texttt{6)} & \texttt{Sched 0.5} & \texttt{Sched \& Slicing 0.5}  \\
\bottomrule
\end{tabular}
\end{adjustbox}
\label{table:default-comp-analysis}
\end{table}

Our results show that \texttt{Sched \& Slicing} and \texttt{Slicing 0.5} favor \gls{embb} the most, with \texttt{Sched \& Slicing 0.5} being the best configuration among the ones considered.
Moreover, slicing is essential to ensure high throughput values (the four top-performing xApps for \gls{embb} include slicing as a control action). We also notice that prioritizing immediate rewards (i.e., $\gamma=0.5$) results in higher throughput values if compared to xApps embedding agents trained to maximize long-term rewards. This design option, when combined with a bigger action space (e.g., scheduling \& slicing) ultimately yields a higher throughput.

For the \gls{mmtc} slice, the \texttt{Slicing 0.99} xApp always yields the best performance. However, we notice that  
\texttt{Sched \& Slicing 0.5}, which is the best-performing xApp for \gls{embb}, yields the worst performance for \gls{mmtc}. 
Although a larger action space and a short-term reward design is ideal for \gls{embb} (e.g., \texttt{Sched \& Slicing 0.5}), we notice that this performance gain comes at the expense of the \gls{mmtc} slice.
Indeed, in Figs.~\ref{fig:Figure4d} and \ref{fig:Figure4e}, we observe that the higher the \gls{embb} performance, the lower the \gls{mmtc}'s. This is clearly illustrated when we compare the \blockquote{best} per-slice policies, respectively \texttt{Sched \& Slicing 0.5} (\gls{embb}) and \texttt{Slicing 0.99} (\gls{mmtc}). The former delivers the highest reported \gls{embb} throughput ($4.168$\:Mbps) but the lowest number of \gls{mmtc} packets ($16$\:packets), while the latter delivers the highest number of \gls{mmtc} packets ($37$\:packets) and one of the lowest measured \gls{embb} throughput values (i.e., $3.636$\:Mbps). 

Hence, \gls{embb}-\gls{mmtc} slices indicate a competitive behavior, since we cannot optimally satisfy both of them without loss in their respective rewards, as they compete for the amount of packets required for transmission. Our results show  that, in general, controlling scheduling only is not ideal as it strongly penalizes \gls{embb} performance with a modest improvement in terms of number of transmitted \gls{mmtc} packets.



\subsection{Impact of Hierarchical Decision-Making}\label{Section IV-B}
\begin{figure*}[t!]
\centering
\subfigure[\gls{embb} Throughput]{\includegraphics[height=2.85cm]{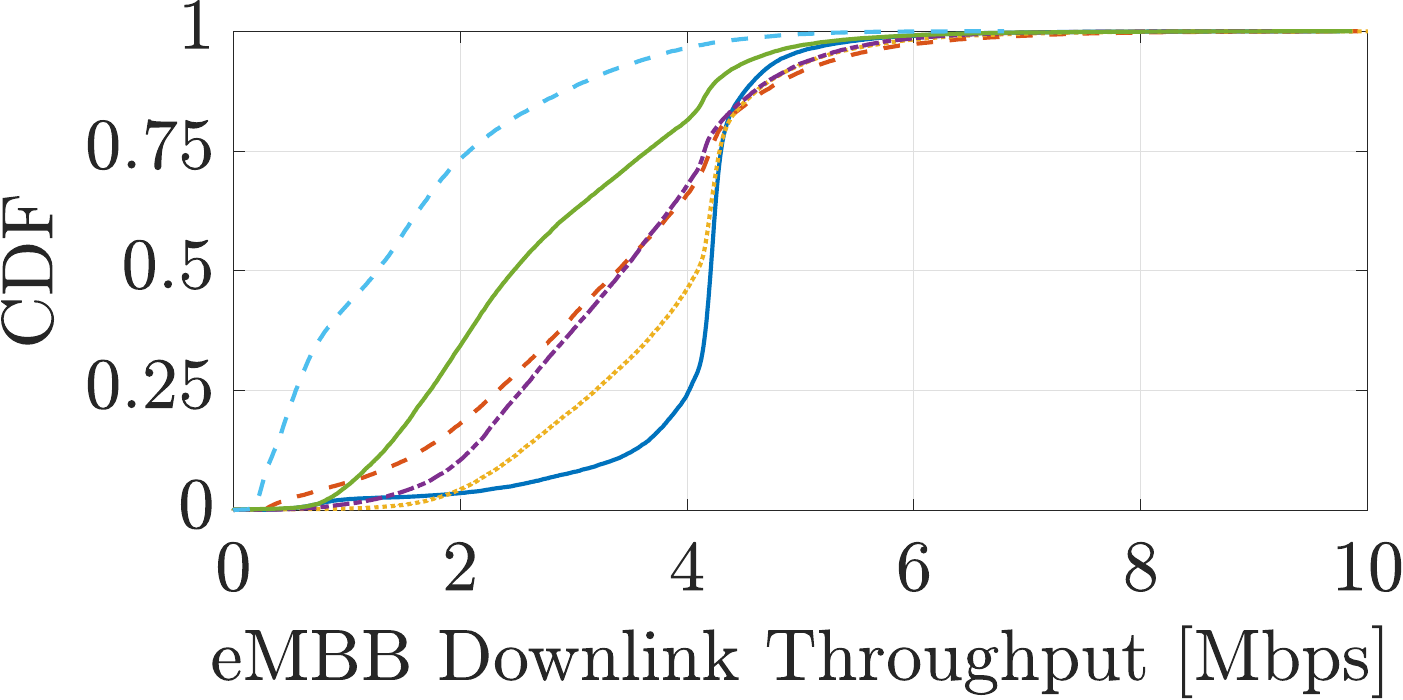}
\label{Figure4a3}}
\hfil
\subfigure[\gls{mmtc} Packets]{\includegraphics[height=2.85cm]{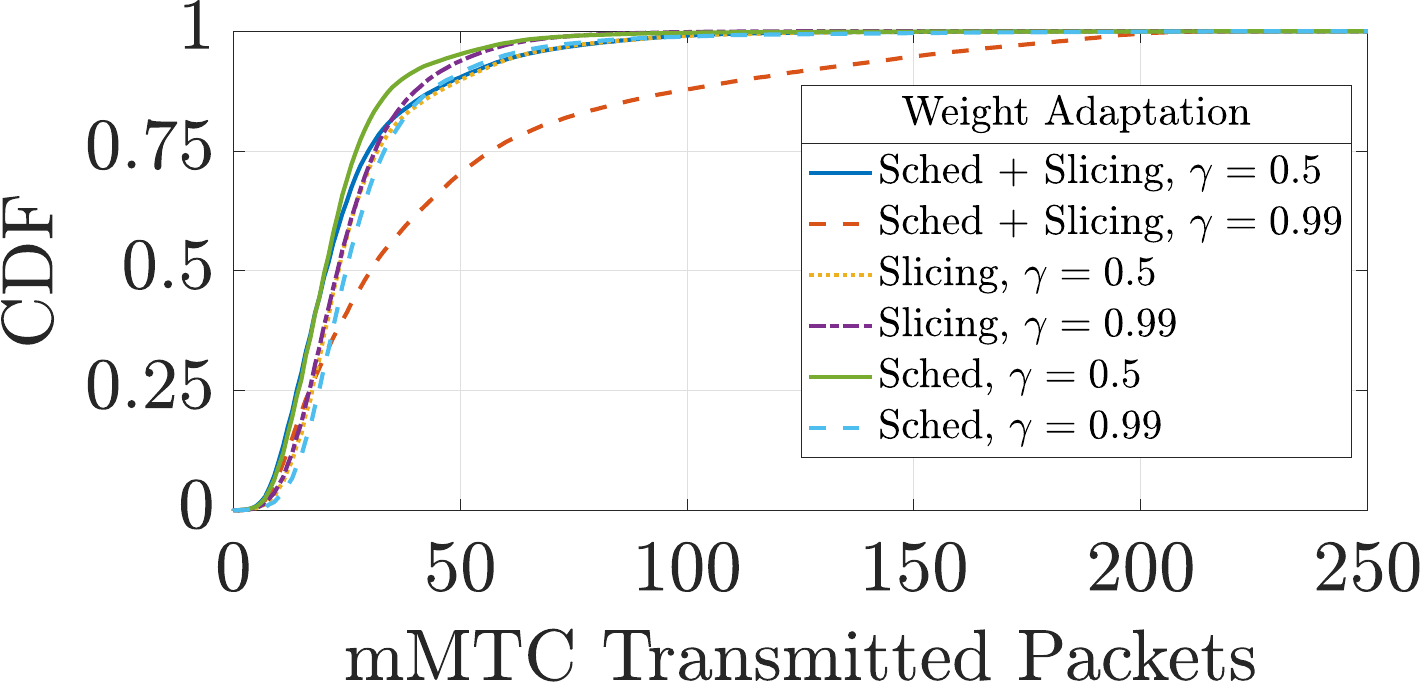}
\label{Figure4b3}}
\hfil
\subfigure[\gls{urllc} Buffer Occupancy]{\includegraphics[height=2.85cm]{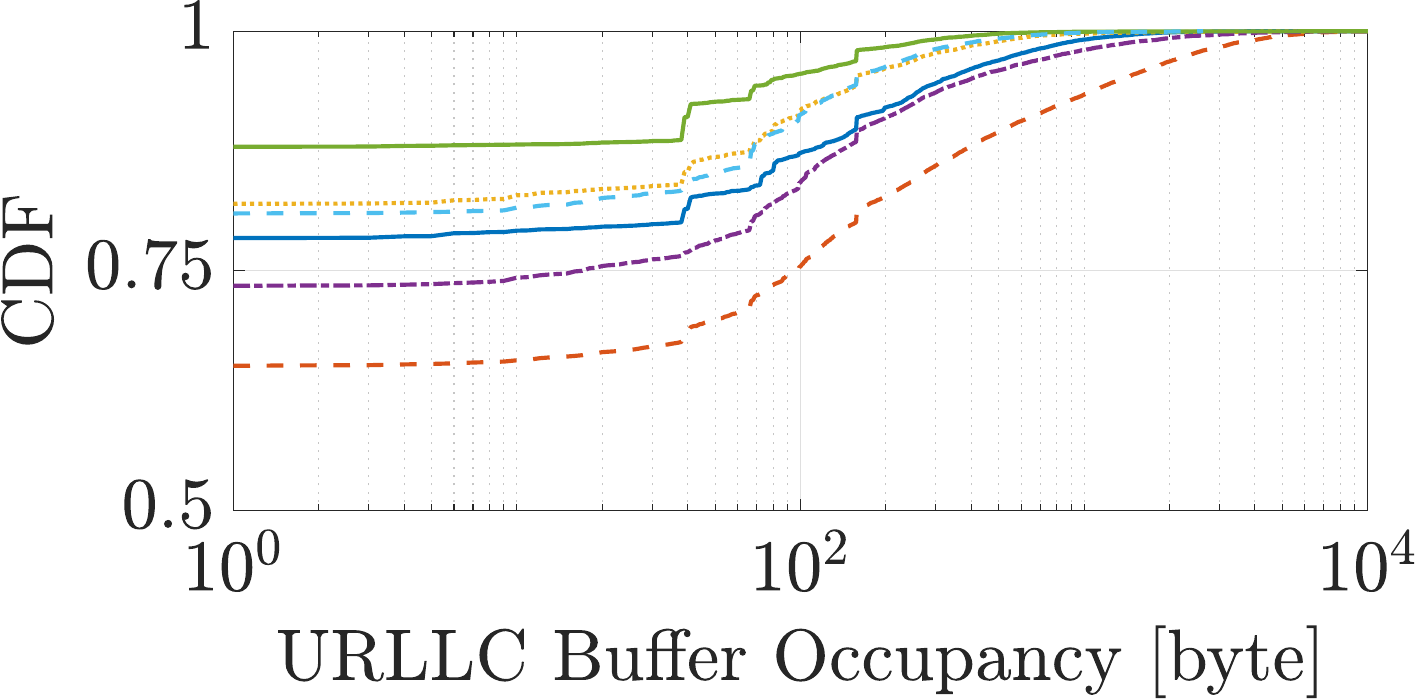}
\label{Figure4c3}}
\setlength\abovecaptionskip{-.02cm}
\caption{Performance evaluation under the \textit{Alternative} weight configuration for different actions spaces and discount factors.}
\label{Figure4-3a}
\vspace{-0.55cm}
\end{figure*}

\begin{figure}[t!]
\centering
\subfigure[\gls{embb} DL Throughput]{\includegraphics[width=1.6in]{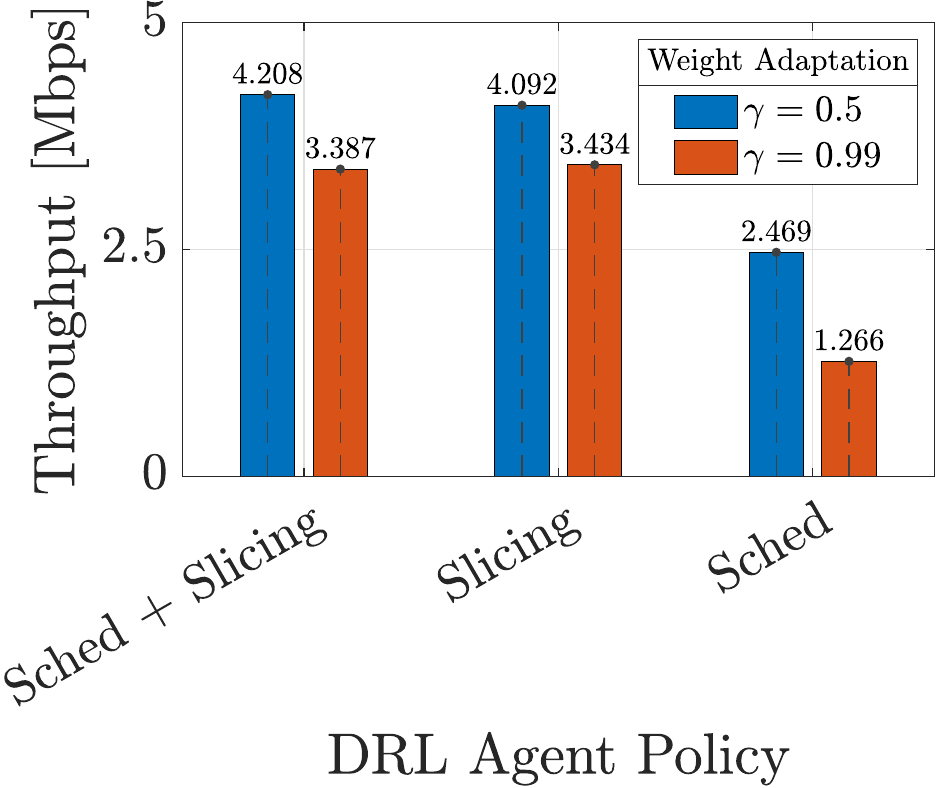}
\label{Figure4d3}}
\hfil
\subfigure[\gls{mmtc}  Packets]{\includegraphics[width=1.6in]{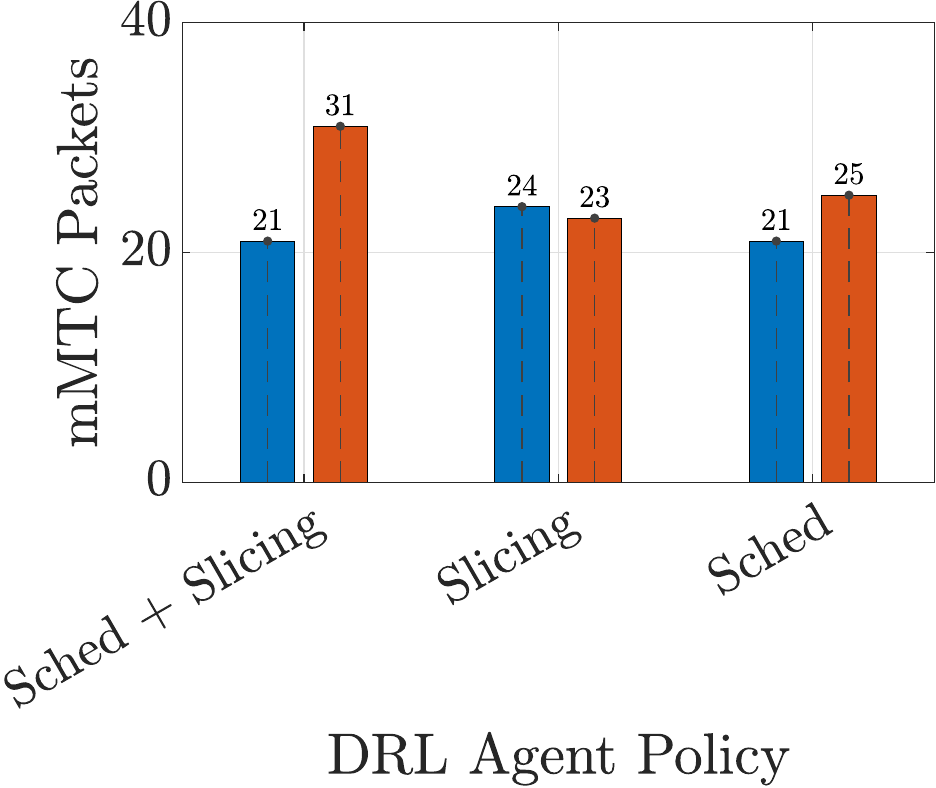}
\label{Figure4e3}}
\setlength\abovecaptionskip{-.02cm}
\caption{Median values under the \textit{Alternative} weight configuration for different actions spaces and discount factors.}
\label{Figure4-3b}
\vspace{-0.45cm}
\end{figure}

In this analysis, we
evaluate the effectiveness of making disjoint decisions to control scheduling and slicing policies.
We select the best performing single-action xApps from Table~\ref{table:default-comp-analysis}, i.e., \texttt{Slicing 0.5} and \texttt{Sched 0.99}, and we compare their execution at different timescales. The former, provides a good balance in terms of \gls{embb} throughput ($\sim 4$ Mbps) and number of \gls{mmtc} packets, while the latter, provides the best performance for the \gls{mmtc} slice. With this design choice, we expect to maintain high performance for both \gls{embb} and \gls{mmtc}. 

We consider four setups, summarized in Table~\ref{table:hier-setups}. Each entry describes how frequently the \gls{bs} reports \glspl{kpm} to the \gls{ric}. For instance, in Setup~1, the xApp for slicing control receives data from the \gls{bs} every $1$\:s, while the scheduling agent receives the respective metrics every $10$\:s. Despite taking into account \gls{ran} telemetry reported every $1, 5$ or $10$\:s, the \gls{drl} decision-making process and the enforcement of a control policy on the \gls{bs} occur within a granularity of sub-milliseconds, and hence the intelligent control loops are still in compliance with the timescale requirements of the near-real-time \gls{ric}.

\begin{table}[ht]
\vspace{-0.1cm}
\centering
\small
\setlength\abovecaptionskip{-.1cm}
\caption{Hierarchical Reporting Setup}
\begin{adjustbox}{width=0.65\linewidth}
\begin{tabular}{@{}c@{}c@{}c@{}}
\toprule
\multicolumn{1}{c}{\textbf{Setup ID}} & \multicolumn{1}{c}{\textbf{\texttt{Slicing 0.5}}} & \multicolumn{1}{c}{\textbf{\texttt{Sched 0.99}}} \\ \midrule
\textbf{1} & $1$\:s & $10$\:s  \\
\textbf{2} & $1$\:s & $5$\:s  \\
\textbf{3} & $10$\:s & $1$\:s \\
\textbf{4} & $5$\:s & $1$ \:s \\
\bottomrule
\end{tabular}
\end{adjustbox}
\label{table:hier-setups}
\vspace{-0.15cm}
\end{table}

Results of this analysis are presented in Figs.~\ref{Figure4-2a} and~\ref{Figure4-2b}. From Fig.~\ref{Figure4a2}, \textit{Setup 3} delivers the best \gls{embb} performance, \textit{Setups} $1$ and $2$ perform almost equally, while \textit{Setup 4} performs the worst. For \gls{mmtc}, in Fig.~\ref{Figure4b2} we notice that all combinations perform similarly and deliver approximately $26$ packets, with \textit{Setup 1} and \textit{Setup 4} delivering an additional packet. From Fig.~\ref{Figure4c2}, we notice that all setups deliver the same performance for the \gls{urllc} slice and, despite not being reported in the figures, they all yield a median buffer occupancy of $0$\:byte, i.e., they maintain an empty buffer to ensure low latency values. In Fig.~\ref{Figure4d2}, we notice that \textit{Setups} $2$ and $3$ deliver the highest \gls{embb} throughput. In Fig.~\ref{Figure4e2}, instead, we notice that \textit{Setups} $1$ and $4$ deliver the highest number of transmitted \gls{mmtc} packets. 

Our findings on hierarchical control verify \gls{embb}'s and \gls{mmtc}'s competitive behavior for individual reward maximization. 
Our results show that the rewards of \gls{embb} and \gls{mmtc} slices are competing with one another, as the best configuration for \gls{embb} corresponds to the worst configuration for \gls{mmtc}, and vice versa.  
Among all considered configurations, \textit{Setup 3} offers the best trade-off, as it delivers the highest throughput at the expense of a single \gls{mmtc} packet less being transmitted. 

\noindent
\subsection{Impact of Weight Configuration}\label{Section IV-C}
In this study, we consider different weight configurations to compute the cumulative average reward function in Eq.~\eqref{eq:weighted_reward}. The considered configurations are reported in Table \ref{table:weight-confs-list}. The \textit{Alternative} weight configuration is computed by using the weights in Table
\ref{table:weight-design}, where $A, B, C, \alpha_{eMBB}, \beta_{mMTC}$, and $\gamma_{URLLC}$ are used to both scale and prioritize certain slices. Specifically, $A, B, C$ are used to scale the individual weights according to statistical information of corresponding \glspl{kpm}. For example, $A, B, C$ can represent either the average, minimum or maximum values reported \gls{kpm} per slice so as to scale the weight according to the dynamic range of the corresponding \gls{kpm}. Similarly, $\alpha_{eMBB}, \beta_{mMTC}$, and $\gamma_{URLLC}$ can be used to give priority to one slice or the other.

\begin{table}[tb]
\centering
\small
\setlength\abovecaptionskip{-.1cm}
\caption{Weight Design}
\begin{adjustbox}{width=0.65\linewidth}
\begin{tabular}{ccc}
\toprule
\textbf{$w_{eMBB}$} & \textbf{$w_{mMTC}$} & \textbf{$w_{URLLC}$} \\
\midrule
$\alpha_{eMBB}\cdot\dfrac{1}{A}$ & $\beta_{mMTC}\cdot\dfrac{1}{B}$ & $\gamma_{URLLC}\cdot \left( -\dfrac{1}{C} \right)$ \\
\bottomrule
\end{tabular}
\end{adjustbox}
\label{table:weight-design}
\vspace{-0.3cm}
\end{table}

We set $\alpha_{eMBB}=1000$, $\beta_{mMTC}=456$ and $\gamma_{URLLC}=1$. As a reference for $A$, $B$ and $C$, we choose the historically maximum reported \gls{kpm} values for each slice, i.e., $A=13.88$ Mbps, $B=304$, and $C=20186$ byte.
 
Based on these steps, we derive their respective weights $w_{eMBB}$, $w_{mMTC}$, $w_{URLLC}$. For example, the weight of \gls{mmtc} can be computed as $w_{mMTC} = \beta_{mMTC}\cdot\dfrac{1}{B} = 456/304 = 1.5$, as reported in the \textit{Alternative} configuration in Table \ref{table:weight-confs-list}.
The goal of comparing the two \textit{Default} and \textit{Alternative} weight configurations is to explore and understand the dynamics between \gls{mmtc} and \gls{embb} and the overall impact on the network performance. Specifically, since previous results have shown that the \gls{mmtc} can be penalized by the \gls{embb} slice, with the \textit{Alternative configuration} we aim at giving the former a weight that is $6\times$ larger than the \textit{Default} configuration.

Results for the \textit{Alternative} configuration are reported in Figs. \ref{Figure4-3a} and \ref{Figure4-3b}.
In Fig. \ref{Figure4a3}, \texttt{Sched \& Slicing 0.5} delivers the best \gls{embb} performance. Similarly to the results presented in Section \ref{Section IV-A}, \texttt{Scheduling \& Slicing 0.5} and \texttt{Slicing 0.5} are the best choices, with short-term reward design being ideal for \gls{embb}. In Fig. \ref{Figure4b3}, the \textit{Alternative} weight configuration results in \texttt{Scheduling \& Slicing 0.99} being the best \gls{mmtc} choice and long-term rewards are better for \gls{mmtc} users. For \gls{urllc}, all policies perform well, with \texttt{Scheduling \& Slicing 0.5} performing slightly better compared to \texttt{Scheduling \& Slicing 0.99}.

Figs.~\ref{Figure4d3} and~\ref{Figure4e3}, confirm that controlling scheduling alone does not improve performance in general. Similarly to our previous analysis, a high \gls{embb} performance (i.e., \texttt{Sched \& Slicing 0.5}) results in a degraded \gls{mmtc} performance.
However, 
if compared with the \texttt{Default}, the \texttt{Alternative} weight configuration achieves a $31.25\%$ increase for \gls{mmtc}, with the same equally good \gls{urllc} performance and a $1\%$ throughput increase for \gls{embb} users. 

In Table~\ref{table:design-catalogue} we summarize the design options that deliver good overall performance. Table~\ref{table:final-designs} indicates \gls{embb} and \gls{mmtc}'s dynamic and competitive relation. Option $2$ brings balance, in terms of throughput and transmitted packets, Option $1$ favors \gls{embb}, and Option $4$ boosts \gls{mmtc} but with a significant decrease in the \gls{qos} of the \gls{embb} slice. 

\begin{table}[htb]
\centering
\small
\setlength\abovecaptionskip{-.1cm}
\caption{Design Options Catalog}
\begin{adjustbox}{width=0.85\linewidth}
\begin{tabular}{@{}l@{\hspace{4.2 mm}}l@{\hspace{4.2 mm}}l@{\hspace{4.2 mm}}l@{}}
\toprule
$\textbf{Option 1}$ & $\texttt{Sched \& Slicing 0.5 - Alternative}$ \\
$\textbf{Option 2}$ & $\texttt{Slicing 0.5 - Default}$ \\
$\textbf{Option 3}$ & $\texttt{Hierarchical Control - Setup 1}$ \\
$\textbf{Option 4}$ & $\texttt{Slicing 0.99 - Default}$ \\
\bottomrule
\end{tabular}
\end{adjustbox}
\label{table:design-catalogue}
\vspace{-0.5cm}
\end{table}

\begin{table}[htb]
\centering
\small
\setlength\abovecaptionskip{-.1cm}
\caption{Design Options}
\begin{adjustbox}{width=0.85\linewidth}
\begin{tabular}
{@{}l@{\hspace{0.1mm}}c@{\hspace{0.1mm}}c@{\hspace{0.1mm}}c@{}}
\toprule
\multicolumn{1}{c}{\textbf{}} & \multicolumn{1}{c}{\textbf{eMBB} [Mbps]} & \multicolumn{1}{c}{\textbf{mMTC} [packet]} & \multicolumn{1}{c}{\textbf{URLLC} [byte]} \\
\midrule
$\textbf{Option 1}$ & $4.208$& $21$ & $0$ \\
$\textbf{Option 2}$ & $4.114$ & $26$ & $0$ \\
$\textbf{Option 3}$ & $3.804$ & $27$ & $0$ \\
$\textbf{Option 4}$ & $3.636$ & $37$ & $0$ \\
\bottomrule
\end{tabular}
\end{adjustbox}
\label{table:final-designs}
\end{table}


\section{Conclusions and Future Work} \label{Section V}

In this paper, we investigated the impact of \gls{drl} design choices on the performance of an Open \gls{ran} system controlled by xApps embedding \gls{drl} agents that make decisions in near-real-time to compute efficient slicing and scheduling control policies.
We
benchmarked 12~xApps trained using \gls{drl} agents with different actions spaces, reward design and decision-making timescales.
Our experimental results show that network slices with similar objectives (e.g., maximizing throughput and number of transmitted packets) might result in a competitive behavior that can be mitigated using proper weight and reward configurations.
Our results point the need for either a) slice-specific xApp design, or b) joint optimization on the premise of xApp conflict avoidance. Part of our current and future work focuses on these directions, with additional testing under diverse \gls{cqi} conditions, mobility patterns and dynamically changing traffic load. Optimal weight design with respect to the size of the action space is also part of our ongoing work. 

\balance
\bibliographystyle{IEEEtran}
\bibliography{IEEEabrv,ref}
\end{document}

%% file: acronyms.tex
\newacronym{3gpp}{3GPP}{3rd Generation Partnership Project}
\newacronym{4g}{4G}{4th generation}
\newacronym{5g}{5G}{5th generation}
\newacronym{6g}{6G}{6th generation}
\newacronym{5gc}{5GC}{5G Core}
\newacronym{adc}{ADC}{Analog to Digital Converter}
\newacronym{aerpaw}{AERPAW}{Aerial Experimentation and Research Platform for Advanced Wireless}
\newacronym{ai}{AI}{Artificial Intelligence}
\newacronym{aimd}{AIMD}{Additive Increase Multiplicative Decrease}
\newacronym{am}{AM}{Acknowledged Mode}
\newacronym{amc}{AMC}{Adaptive Modulation and Coding}
\newacronym{amf}{AMF}{Access and Mobility Management Function}
\newacronym{aops}{AOPS}{Adaptive Order Prediction Scheduling}
\newacronym{api}{API}{Application Programming Interface}
\newacronym{xapp}{xApp}{Intelligent Application}
\newacronym{apn}{APN}{Access Point Name}
\newacronym{ap}{AP}{Application Protocol}
\newacronym{aqm}{AQM}{Active Queue Management}
\newacronym{ausf}{AUSF}{Authentication Server Function}
\newacronym{avc}{AVC}{Advanced Video Coding}
\newacronym{awgn}{AGWN}{Additive White Gaussian Noise}
\newacronym{balia}{BALIA}{Balanced Link Adaptation Algorithm}
\newacronym{bbu}{BBU}{Base Band Unit}
\newacronym{bdp}{BDP}{Bandwidth-Delay Product}
\newacronym{ber}{BER}{Bit Error Rate}
\newacronym{bf}{BF}{Beamforming}
\newacronym{bler}{BLER}{Block Error Rate}
\newacronym{brr}{BRR}{Bayesian Ridge Regressor}
\newacronym{bs}{BS}{Base Station}
\newacronym{bsr}{BSR}{Buffer Status Report}
\newacronym{bss}{BSS}{Business Support System}
\newacronym{ca}{CA}{Carrier Aggregation}
\newacronym{caas}{CaaS}{Connectivity-as-a-Service}
\newacronym{cb}{CB}{Code Block}
\newacronym{cc}{CC}{Congestion Control}
\newacronym{ccid}{CCID}{Congestion Control ID}
\newacronym{cco}{CC}{Carrier Component}
\newacronym{cdd}{CDD}{Cyclic Delay Diversity}
\newacronym{cdf}{CDF}{Cumulative Distribution Function}
\newacronym{cdn}{CDN}{Content Distribution Network}
\newacronym{cn}{CN}{Core Network}
\newacronym{codel}{CoDel}{Controlled Delay Management}
\newacronym{comac}{COMAC}{Converged Multi-Access and Core}
\newacronym{cord}{CORD}{Central Office Re-architected as a Datacenter}
\newacronym{cornet}{CORNET}{COgnitive Radio NETwork}
\newacronym{cosmos}{COSMOS}{Cloud Enhanced Open Software Defined Mobile Wireless Testbed for City-Scale Deployment}
\newacronym{cots}{COTS}{Commercial Off-the-Shelf}
\newacronym{cp}{CP}{Control Plane}
\newacronym{cpu}{CPU}{Central Processing Unit}
\newacronym{cqi}{CQI}{Channel Quality Information}
\newacronym{cr}{CR}{Cognitive Radio}
\newacronym{cran}{CRAN}{Cloud \gls{ran}}
\newacronym{crs}{CRS}{Cell Reference Signal}
\newacronym{csi}{CSI}{Channel State Information}
\newacronym{csirs}{CSI-RS}{Channel State Information - Reference Signal}
\newacronym{cu}{CU}{Central Unit}
\newacronym{d2tcp}{D$^2$TCP}{Deadline-aware Data center TCP}
\newacronym{d3}{D$^3$}{Deadline-Driven Delivery}
\newacronym{dac}{DAC}{Digital to Analog Converter}
\newacronym{dag}{DAG}{Directed Acyclic Graph}
\newacronym{das}{DAS}{Distributed Antenna System}
\newacronym{dash}{DASH}{Dynamic Adaptive Streaming over HTTP}
\newacronym{dc}{DC}{Dual Connectivity}
\newacronym{dccp}{DCCP}{Datagram Congestion Control Protocol}
\newacronym{dce}{DCE}{Direct Code Execution}
\newacronym{dci}{DCI}{Downlink Control Information}
\newacronym{dctcp}{DCTCP}{Data Center TCP}
\newacronym{dl}{DL}{Downlink}
\newacronym{dmr}{DMR}{Deadline Miss Ratio}
\newacronym{dmrs}{DMRS}{DeModulation Reference Signal}
\newacronym{drlcc}{DRL-CC}{Deep Reinforcement Learning Congestion Control}
\newacronym{drs}{DRS}{Discovery Reference Signal}
\newacronym{du}{DU}{Distributed Unit}
\newacronym{e2e}{E2E}{end-to-end}
\newacronym{ecaas}{ECaaS}{Edge-Cloud-as-a-Service}
\newacronym{ecn}{ECN}{Explicit Congestion Notification}
\newacronym{edf}{EDF}{Earliest Deadline First}
\newacronym{embb}{eMBB}{Enhanced Mobile Broadband}
\newacronym{empower}{EMPOWER}{EMpowering transatlantic PlatfOrms for advanced WirEless Research}
\newacronym{enb}{eNB}{evolved Node Base}
\newacronym{endc}{EN-DC}{E-UTRAN-\gls{nr} \gls{dc}}
\newacronym{epc}{EPC}{Evolved Packet Core}
\newacronym{eps}{EPS}{Evolved Packet System}
\newacronym{es}{ES}{Edge Server}
\newacronym{etsi}{ETSI}{European Telecommunications Standards Institute}
\newacronym[firstplural=Estimated Times of Arrival (ETAs)]{eta}{ETA}{Estimated Time of Arrival}
\newacronym{eutran}{E-UTRAN}{Evolved Universal Terrestrial Access Network}
\newacronym{faas}{FaaS}{Function-as-a-Service}
\newacronym{fapi}{FAPI}{Functional Application Platform Interface}
\newacronym{fdd}{FDD}{Frequency Division Duplexing}
\newacronym{fdm}{FDM}{Frequency Division Multiplexing}
\newacronym{fdma}{FDMA}{Frequency Division Multiple Access}
\newacronym{fed4fire}{FED4FIRE+}{Federation 4 Future Internet Research and Experimentation Plus}
\newacronym{fir}{FIR}{Finite Impulse Response}
\newacronym{fit}{FIT}{Future \acrlong{iot}}
\newacronym{fpga}{FPGA}{Field Programmable Gate Array}
\newacronym{fr2}{FR2}{Frequency Range 2}
\newacronym{fs}{FS}{Fast Switching}
\newacronym{fscc}{FSCC}{Flow Sharing Congestion Control}
\newacronym{ftp}{FTP}{File Transfer Protocol}
\newacronym{fw}{FW}{Flow Window}
\newacronym{ge}{GE}{Gaussian Elimination}
\newacronym{gnb}{gNB}{Next Generation Node Base}
\newacronym{nextg}{NextG}{Next Generation}
\newacronym{gop}{GOP}{Group of Pictures}
\newacronym{gpr}{GPR}{Gaussian Process Regressor}
\newacronym{gpu}{GPU}{Graphics Processing Unit}
\newacronym{gtp}{GTP}{GPRS Tunneling Protocol}
\newacronym{gtpc}{GTP-C}{GPRS Tunnelling Protocol Control Plane}
\newacronym{gtpu}{GTP-U}{GPRS Tunnelling Protocol User Plane}
\newacronym{gtpv2c}{GTPv2-C}{\gls{gtp} v2 - Control}
\newacronym{gw}{GW}{Gateway}
\newacronym{harq}{HARQ}{Hybrid Automatic Repeat reQuest}
\newacronym{hetnet}{HetNet}{Heterogeneous Network}
\newacronym{hh}{HH}{Hard Handover}
\newacronym{hol}{HOL}{Head-of-Line}
\newacronym{hqf}{HQF}{Highest-quality-first}
\newacronym{hss}{HSS}{Home Subscription Server}
\newacronym{http}{HTTP}{HyperText Transfer Protocol}
\newacronym{ia}{IA}{Initial Access}
\newacronym{iab}{IAB}{Integrated Access and Backhaul}
\newacronym{ic}{IC}{Incident Command}
\newacronym{ietf}{IETF}{Internet Engineering Task Force}
\newacronym{imsi}{IMSI}{International Mobile Subscriber Identity}
\newacronym{imt}{IMT}{International Mobile Telecommunication}
\newacronym{iot}{IoT}{Internet of Things}
\newacronym{ip}{IP}{Internet Protocol}
\newacronym{itu}{ITU}{International Telecommunication Union}
\newacronym{kpi}{KPI}{Key Performance Indicator}
\newacronym{kpm}{KPM}{Key Performance Measurement}
\newacronym{kvm}{KVM}{Kernel-based Virtual Machine}
\newacronym{los}{LOS}{Line-of-Sight}
\newacronym{lsm}{LSM}{Link-to-System Mapping}
\newacronym{lstm}{LSTM}{Long Short Term Memory}
\newacronym{lte}{LTE}{Long Term Evolution}
\newacronym{lxc}{LXC}{Linux Container}
\newacronym{m2m}{M2M}{Machine to Machine}
\newacronym{mac}{MAC}{Medium Access Control}
\newacronym{manet}{MANET}{Mobile Ad Hoc Network}
\newacronym{mano}{MANO}{Management and Orchestration}
\newacronym{mc}{MC}{Multi-Connectivity}
\newacronym{mcc}{MCC}{Mobile Cloud Computing}
\newacronym{mchem}{MCHEM}{Massive Channel Emulator}
\newacronym{mcs}{MCS}{Modulation and Coding Scheme}
\newacronym{mec}{MEC}{Multi-access Edge Computing}
\newacronym{mec2}{MEC}{Mobile Edge Cloud}
\newacronym{mfc}{MFC}{Mobile Fog Computing}
\newacronym{mgen}{MGEN}{Multi-Generator}
\newacronym{mi}{MI}{Mutual Information}
\newacronym{mib}{MIB}{Master Information Block}
\newacronym{miesm}{MIESM}{Mutual Information Based Effective SINR}
\newacronym{mimo}{MIMO}{Multiple Input, Multiple Output}
\newacronym{ml}{ML}{Machine Learning}
\newacronym{mlr}{MLR}{Maximum-local-rate}
\newacronym[plural=\gls{mme}s,firstplural=Mobility Management Entities (MMEs)]{mme}{MME}{Mobility Management Entity}
\newacronym{mmtc}{mMTC}{Massive Machine-Type Communications}
\newacronym{mmwave}{mmWave}{millimeter wave}
\newacronym{mpdccp}{MP-DCCP}{Multipath Datagram Congestion Control Protocol}
\newacronym{mptcp}{MPTCP}{Multipath TCP}
\newacronym{mr}{MR}{Maximum Rate}
\newacronym{mrdc}{MR-DC}{Multi \gls{rat} \gls{dc}}
\newacronym{mse}{MSE}{Mean Square Error}
\newacronym{mss}{MSS}{Maximum Segment Size}
\newacronym{mt}{MT}{Mobile Termination}
\newacronym{mtd}{MTD}{Machine-Type Device}
\newacronym{mtu}{MTU}{Maximum Transmission Unit}
\newacronym{mumimo}{MU-MIMO}{Multi-user \gls{mimo}}
\newacronym{mvno}{MVNO}{Mobile Virtual Network Operator}
\newacronym{nalu}{NALU}{Network Abstraction Layer Unit}
\newacronym{nas}{NAS}{Non-Access Stratum}
\newacronym{nbiot}{NB-IoT}{Narrow Band IoT}
\newacronym{nfv}{NFV}{Network Function Virtualization}
\newacronym{nfvi}{NFVI}{Network Function Virtualization Infrastructure}
\newacronym{nic}{NIC}{Network Interface Card}
\newacronym{nlos}{NLOS}{Non-Line-of-Sight}
\newacronym{now}{NOW}{Non Overlapping Window}
\newacronym{nsm}{NSM}{Network Service Mesh}
\newacronym[type=hidden]{nr}{NR}{New Radio}
\newacronym{nrf}{NRF}{Network Repository Function}
\newacronym{nsa}{NSA}{Non Stand Alone}
\newacronym{nse}{NSE}{Network Slicing Engine}
\newacronym{nssf}{NSSF}{Network Slice Selection Function}
\newacronym{o2i}{O2I}{Outdoor to Indoor}
\newacronym{oai}{OAI}{OpenAirInterface}
\newacronym{oaicn}{OAI-CN}{\gls{oai} \acrlong{cn}}
\newacronym{oairan}{OAI-RAN}{\acrlong{oai} \acrlong{ran}}
\newacronym{oam}{OAM}{Operations, Administration and Maintenance}
\newacronym{ofdm}{OFDM}{Orthogonal Frequency Division Multiplexing}
\newacronym{olia}{OLIA}{Opportunistic Linked Increase Algorithm}
\newacronym{omec}{OMEC}{Open Mobile Evolved Core}
\newacronym{onap}{ONAP}{Open Network Automation Platform}
\newacronym{onf}{ONF}{Open Networking Foundation}
\newacronym{onos}{ONOS}{Open Networking Operating System}
\newacronym{oom}{OOM}{\gls{onap} Operations Manager}
\newacronym{opnfv}{OPNFV}{Open Platform for \gls{nfv}}
\newacronym{orbit}{ORBIT}{Open-Access Research Testbed for Next-Generation Wireless Networks}
\newacronym{os}{OS}{Operating System}
\newacronym{oss}{OSS}{Operations Support System}
\newacronym{pa}{PA}{Position-aware}
\newacronym{pase}{PASE}{Prioritization, Arbitration, and Self-adjusting Endpoints}
\newacronym{pawr}{PAWR}{Platforms for Advanced Wireless Research}
\newacronym{pbch}{PBCH}{Physical Broadcast Channel}
\newacronym{pcef}{PCEF}{Policy and Charging Enforcement Function}
\newacronym{pcfich}{PCFICH}{Physical Control Format Indicator Channel}
\newacronym{pcrf}{PCRF}{Policy and Charging Rules Function}
\newacronym{pdcch}{PDCCH}{Physical Downlink Control Channel}
\newacronym{pdcp}{PDCP}{Packet Data Convergence Protocol}
\newacronym{pdsch}{PDSCH}{Physical Downlink Shared Channel}
\newacronym{pdu}{PDU}{Packet Data Unit}
\newacronym{pf}{PF}{Proportionally Fair}
\newacronym{pgw}{PGW}{Packet Gateway}
\newacronym{phich}{PHICH}{Physical Hybrid ARQ Indicator Channel}
\newacronym{phy}{PHY}{Physical}
\newacronym{pmch}{PMCH}{Physical Multicast Channel}
\newacronym{pmi}{PMI}{Precoding Matrix Indicators}
\newacronym{powder}{POWDER}{Platform for Open Wireless Data-driven Experimental Research}
\newacronym{ppo}{PPO}{Proximal Policy Optimization}
\newacronym{ppp}{PPP}{Poisson Point Process}
\newacronym{prach}{PRACH}{Physical Random Access Channel}
\newacronym{prb}{PRB}{Physical Resource Block}
\newacronym{psnr}{PSNR}{Peak Signal to Noise Ratio}
\newacronym{pss}{PSS}{Primary Synchronization Signal}
\newacronym{pucch}{PUCCH}{Physical Uplink Control Channel}
\newacronym{pusch}{PUSCH}{Physical Uplink Shared Channel}
\newacronym{qam}{QAM}{Quadrature Amplitude Modulation}
\newacronym{qci}{QCI}{\gls{qos} Class Identifier}
\newacronym{qoe}{QoE}{Quality of Experience}
\newacronym{qos}{QoS}{Quality of Service}
\newacronym{quic}{QUIC}{Quick UDP Internet Connections}
\newacronym{rach}{RACH}{Random Access Channel}
\newacronym{ran}{RAN}{Radio Access Network}
\newacronym[firstplural=end to endcess Technologies (RATs)]{rat}{RAT}{end to endcess Technology}
\newacronym{rcn}{RCN}{Research Coordination Network}
\newacronym{rec}{REC}{Radio Edge Cloud}
\newacronym{ra}{RA}{Resource Allocation}
\newacronym{red}{RED}{Random Early Detection}
\newacronym{renew}{RENEW}{Reconfigurable Eco-system for Next-generation End-to-end Wireless}
\newacronym{rf}{RF}{Radio Frequency}
\newacronym{rfc}{RFC}{Request for Comments}
\newacronym{rfr}{RFR}{Random Forest Regressor}
\newacronym{ric}{RIC}{RAN Intelligent Controller}
\newacronym{rlc}{RLC}{Radio Link Control}
\newacronym{rlf}{RLF}{Radio Link Failure}
\newacronym{rlnc}{RLNC}{Random Linear Network Coding}
\newacronym{rmr}{RMR}{RIC Message Router}
\newacronym{rmse}{RMSE}{Root Mean Squared Error}
\newacronym{rnis}{RNIS}{Radio Network Information Service}
\newacronym{rr}{RR}{Round Robin}
\newacronym{rrc}{RRC}{Radio Resource Control}
\newacronym{rrm}{RRM}{Radio Resource Management}
\newacronym{rru}{RRU}{Remote Radio Unit}
\newacronym{rs}{RS}{Remote Server}
\newacronym{rsrp}{RSRP}{Reference Signal Received Power}
\newacronym{rsrq}{RSRQ}{Reference Signal Received Quality}
\newacronym{rss}{RSS}{Received Signal Strength}
\newacronym{rssi}{RSSI}{Received Signal Strength Indicator}
\newacronym{rtt}{RTT}{Round Trip Time}
\newacronym{ru}{RU}{Radio Unit}
\newacronym{rw}{RW}{Receive Window}
\newacronym{rx}{RX}{Receiver}
\newacronym{s1ap}{S1AP}{S1 Application Protocol}
\newacronym{sa}{SA}{standalone}
\newacronym{sack}{SACK}{Selective Acknowledgment}
\newacronym{sap}{SAP}{Service Access Point}
\newacronym{sc2}{SC2}{Spectrum Collaboration Challenge}
\newacronym{scef}{SCEF}{Service Capability Exposure Function}
\newacronym{sch}{SCH}{Secondary Cell Handover}
\newacronym{scoot}{SCOOT}{Split Cycle Offset Optimization Technique}
\newacronym{sctp}{SCTP}{Stream Control Transmission Protocol}
\newacronym{sdap}{SDAP}{Service Data Adaptation Protocol}
\newacronym{sdk}{SDK}{Software Development Kit}
\newacronym{sdm}{SDM}{Space Division Multiplexing}
\newacronym{sdma}{SDMA}{Spatial Division Multiple Access}
\newacronym{sdn}{SDN}{Software-defined Networking}
\newacronym{sdr}{SDR}{Software-defined Radio}
\newacronym{seba}{SEBA}{SDN-Enabled Broadband Access}
\newacronym{sgsn}{SGSN}{Serving GPRS Support Node}
\newacronym{sgw}{SGW}{Service Gateway}
\newacronym{si}{SI}{Study Item}
\newacronym{sib}{SIB}{Secondary Information Block}
\newacronym{sinr}{SINR}{Signal to Interference plus Noise Ratio}
\newacronym{sip}{SIP}{Session Initiation Protocol}
\newacronym{siso}{SISO}{Single Input, Single Output}
\newacronym{sla}{SLA}{Service Level Agreement}
\newacronym{sm}{SM}{Service Model}
\newacronym{smf}{SMF}{Session Management Function}
\newacronym{smo}{SMO}{Service Management and Orchestration}
\newacronym{sms}{SMS}{Short Message Service}
\newacronym{smsgmsc}{SMS-GMSC}{\gls{sms}-Gateway}
\newacronym{snr}{SNR}{Signal-to-Noise-Ratio}
\newacronym{son}{SON}{Self-Organizing Network}
\newacronym{sptcp}{SPTCP}{Single Path TCP}
\newacronym{srb}{SRB}{Service Radio Bearer}
\newacronym{srn}{SRN}{Standard Radio Node}
\newacronym{srs}{SRS}{Sounding Reference Signal}
\newacronym{ss}{SS}{Synchronization Signal}
\newacronym{sss}{SSS}{Secondary Synchronization Signal}
\newacronym{st}{ST}{Spanning Tree}
\newacronym{svc}{SVC}{Scalable Video Coding}
\newacronym{tb}{TB}{Transport Block}
\newacronym{tcp}{TCP}{Transmission Control Protocol}
\newacronym{tdd}{TDD}{Time Division Duplexing}
\newacronym{tdm}{TDM}{Time Division Multiplexing}
\newacronym{tdma}{TDMA}{Time Division Multiple Access}
\newacronym{tfl}{TfL}{Transport for London}
\newacronym{tfrc}{TFRC}{TCP-Friendly Rate Control}
\newacronym{tft}{TFT}{Traffic Flow Template}
\newacronym{tgen}{TGEN}{Traffic Generator}
\newacronym{tip}{TIP}{Telecom Infra Project}
\newacronym{tm}{TM}{Transparent Mode}
\newacronym{to}{TO}{Telco Operator}
\newacronym{tr}{TR}{Technical Report}
\newacronym{trp}{TRP}{Transmitter Receiver Pair}
\newacronym{ts}{TS}{Technical Specification}
\newacronym{tti}{TTI}{Transmission Time Interval}
\newacronym{ttt}{TTT}{Time-to-Trigger}
\newacronym{tx}{TX}{Transmitter}
\newacronym{uas}{UAS}{Unmanned Aerial System}
\newacronym{uav}{UAV}{Unmanned Aerial Vehicle}
\newacronym{udm}{UDM}{Unified Data Management}
\newacronym{udp}{UDP}{User Datagram Protocol}
\newacronym{udr}{UDR}{Unified Data Repository}
\newacronym{ue}{UE}{User Equipment}
\newacronym{uhd}{UHD}{\gls{usrp} Hardware Driver}
\newacronym{ul}{UL}{Uplink}
\newacronym{um}{UM}{Unacknowledged Mode}
\newacronym{uml}{UML}{Unified Modeling Language}
\newacronym{upa}{UPA}{Uniform Planar Array}
\newacronym{upf}{UPF}{User Plane Function}
\newacronym{urllc}{URLLC}{Ultra Reliable and Low Latency Communications}
\newacronym{usa}{U.S.}{United States}
\newacronym{usim}{USIM}{Universal Subscriber Identity Module}
\newacronym{usrp}{USRP}{Universal Software Radio Peripheral}
\newacronym{utc}{UTC}{Urban Traffic Control}
\newacronym{vim}{VIM}{Virtualization Infrastructure Manager}
\newacronym{vm}{VM}{Virtual Machine}
\newacronym{vnf}{VNF}{Virtual Network Function}
\newacronym{volte}{VoLTE}{Voice over \gls{lte}}
\newacronym{voltha}{VOLTHA}{Virtual OLT HArdware Abstraction}
\newacronym{vr}{VR}{Virtual Reality}
\newacronym{vran}{vRAN}{Virtualized \gls{ran}}
\newacronym{vss}{VSS}{Video Streaming Server}
\newacronym{wbf}{WBF}{Wired Bias Function}
\newacronym{wf}{WF}{Waterfilling}
\newacronym{wlan}{WLAN}{Wireless Local Area Network}
\newacronym{osm}{OSM}{Open Source \gls{nfv} Management and Orchestration}
\newacronym{pnf}{PNF}{Physical Network Function}
\newacronym{drl}{DRL}{Deep Reinforcement Learning}
\newacronym{rl}{RL}{Reinforcement Learning}
\newacronym{mtc}{MTC}{Machine-type Communications}
\newacronym{osc}{OSC}{O-RAN Software Community}
\newacronym{rc}{RC}{RAN Control}
\newacronym{dqn}{DQN}{Deep Q-Network}
\newacronym{v2x}{V2X}{Vehicle-to-everything}
\newacronym{gbsm}{GBSM}{Geometry-Based Stochastic Model}
\newacronym{gbs}{GBSM}{Geometry-Based Stochastic}
\newacronym{quadriga}{QuaDRiGa}{QUAsi Deterministic RadIo channel GenerAtor}
\newacronym{relu}{ReLU}{Rectified Linear Unit} 
\newacronym{mpc}{MPC}{Multipath Component}